\documentclass[12pt]{article}
\usepackage{latexsym,amsmath,amssymb,bm,cite}
\usepackage{graphicx}
\usepackage{color}
\newcommand{\be}{\begin{equation}}
\newcommand{\ee}{\end{equation}}
\newcommand{\bea}{\begin{eqnarray}}
\newcommand{\eea}{\end{eqnarray}}
\newcommand{\bem}{\begin{multline}}
\newcommand{\eem}{\end{multline}}
\newcommand{\beg}{\begin{gather}}
\newcommand{\eeg}{\end{gather}}

\newcommand{\as}{\alpha_s}

 \newcommand{\bal}{{\bar\alpha}}
 \def\eq#1{{Eq.~(\ref{#1})}}
 \def\fig#1{{Fig.~\ref{#1}}}
\newcommand{\am}{\alpha_\mu}

\begin{document}
\title{{\bf Collinear Singularities and Running Coupling  \\[.5cm] Corrections to Gluon Production in CGC \\[1.5cm] }}
\author{
{\bf Yuri V.\ Kovchegov$^1$ and Heribert Weigert$^2$}
\\[1cm] {\it\small $^1$Department of Physics, The Ohio State University}\\ 
{\it\small Columbus, OH 43210, USA}\\[5mm]
{\it\small $^2$Department of Physics, University of Oulu, P.O. Box 3000}\\ 
{\it\small FI-90014 Oulu, Finland}\\[5mm]}

\date{April 2008}

\maketitle

\thispagestyle{empty}

\begin{abstract}
  We analyze the structure of running coupling corrections to the gluon
  production cross section in the projectile--nucleus collisions calculated in
  the Color Glass Condensate (CGC) framework.  We argue that for the gluon
  production cross section (and for gluon transverse momentum spectra and
  multiplicity) the inclusion of running coupling corrections brings in
  collinear singularities due to final state splittings completely unaffected
  by CGC resummations. Hence, despite the saturation/CGC dynamics, the gluon
  production cross section is not infrared-safe. As usual, regularizing the
  singularities requires an infrared cutoff $\Lambda_{\text{coll}}$ that
  defines a resolution scale for gluons.  We specifically show that the cutoff
  enters the gluon production cross section in the argument of the strong
  coupling constant $\as (\Lambda_{\text{coll}}^2)$. We argue that for hadron
  production calculations one should be able to absorb the collinear
  divergence into a fragmentation function.  The singular collinear terms in
  the gluon production cross section are shown not to contribute to the energy
  density $\epsilon$ of the produced matter, which is indeed an
  infrared-finite quantity.
\end{abstract}

\thispagestyle{empty}

\newpage

\setcounter{page}{1}


\section{Introduction}

In the recent years there has been a lot of progress in understanding
running coupling corrections for small-$x$ evolution equations.
Running coupling corrections to the Balitsky-Fadin-Kuraev-Lipatov
(BFKL) \cite{Kuraev:1977fs,Bal-Lip}, the Balitsky-Kovchegov (BK)
\cite{Balitsky:1996ub,
  Balitsky:1997mk,Balitsky:1998ya,Kovchegov:1999yj, Kovchegov:1999ua}
and the Jalilian-Marian--Iancu--McLerran--Weigert--Leonidov--Kovner
(JIMWLK) \cite{Jalilian-Marian:1997jx, Jalilian-Marian:1997gr,
  Jalilian-Marian:1997dw, Jalilian-Marian:1998cb, Kovner:2000pt,
  Weigert:2000gi, Iancu:2000hn,Ferreiro:2001qy} evolution equations
were calculated in
\cite{Balitsky:2006wa,Kovchegov:2006vj,Kovchegov:2006wf,Albacete:2007yr}.
Knowledge of the running coupling corrections to the BK and JIMWLK
equations significantly improved our ability to make precise
predictions for the total cross sections in deep inelastic scattering
(DIS) based on the saturation/Color Glass Condensate (CGC) physics
\cite{Gribov:1984tu,Mueller:1986wy,McLerran:1994vd,McLerran:1993ka,
  McLerran:1993ni,Kovchegov:1996ty,Kovchegov:1997pc,Jalilian-Marian:1997xn,
  Jalilian-Marian:1997jx, Jalilian-Marian:1997gr,
  Jalilian-Marian:1997dw, Jalilian-Marian:1998cb, Kovner:2000pt,
  Weigert:2000gi, Iancu:2000hn,Ferreiro:2001qy,Kovchegov:1999yj,
  Kovchegov:1999ua, Balitsky:1996ub, Balitsky:1997mk,
  Balitsky:1998ya,Iancu:2003xm,Weigert:2005us,Jalilian-Marian:2005jf}.

However, to improve the CGC predictions for other observables, such as
the inclusive gluon or quark production cross sections, it is
important to understand how the running coupling corrections enter the
expressions derived previously for such cross sections
\cite{Kovner:1995ja,Kovner:1995ts,Gyulassy:1997vt,Kovchegov:1997ke,Kovchegov:1998bi,Kopeliovich:1998nw,Dumitru:2001ux,Kovchegov:2001sc,Gelis:2002nn,Kovchegov:2006qn,Kovchegov:2005ur,Dumitru:2001jn,Dumitru:2002qt,Kovner:2001vi,Blaizot:2004wu,Marquet:2004xa,Gelis:2005pt}.
It was originally proposed in \cite{Gribov:1984tu,Gribov:1981ac} that
the gluon production cross section can be described by the
$k_T$-factorization formula
\begin{align}\label{ktfact}
  \frac{d \sigma}{d^2 p \, dy} \, = \, \frac{2 \, \as}{C_F} \,
  \frac{1}{p_T^2} \, \int d^2 q \, \phi_p ({\bm q}, y) \, \phi_t ({\bm p}
  - {\bm q}, Y - y),
\end{align}
where the gluon is produced with the transverse momentum $\bm p$ ($p_T
= |{\bm p}|$), rapidity $y$ in a collision with the total rapidity
interval $Y$. $\phi_p$ and $\phi_t$ are the unintegrated gluon
distribution functions in the projectile and in the target
correspondingly. $\as$ is the strong coupling constant and $C_F =
(N_c^2 -1)/2 N_c$. 

$k_T$-factorization (\ref{ktfact}) has been known to work outside the
saturation region
\cite{Catani:1990eg,Kovner:1995ja,Kovner:1995ts,Gyulassy:1997vt,Kovchegov:1997ke}
(e.g. for proton-proton collisions $pp$). In
\cite{Kovchegov:2001sc,Kharzeev:2003wz,Kovchegov:2005ur} it was shown
that the $k_T$-factorization formula (\ref{ktfact}) holds for gluon
production in proton-nucleus ($pA$) collisions and in DIS both in the
quasi-classical limit of multiple rescatterings
\cite{McLerran:1994vd,McLerran:1993ka,
  McLerran:1993ni,Kovchegov:1996ty} and after including quantum
BK/JIMWLK evolution. This means that \eq{ktfact} works in the
kinematic region inside the saturation region for the target and
outside the saturation region for the projectile. (A proton/projectile
can be defined as a nucleus with a much smaller saturation scale than
in the target nucleus.) It is at present not clear whether \eq{ktfact}
correctly describes gluon production in nucleus-nucleus collisions,
i.e., in the saturation region for both the target and the projectile.

In \cite{Kovchegov:2001sc} the following relation was also derived
between the forward amplitude of a dipole of transverse size $\bm x$
scattering on a nucleus at impact parameter $\bm b$ and rapidity $y$,
$N ({\bm x}, {\bm b}, y)$, and the unintegrated gluon distribution
function $\phi ({\bm q}, y)$ of the same nucleus:
\begin{align}\label{phi}
  \phi ({\bm q}, y) \, = \, \frac{C_F}{\as \, (2 \, \pi)^3} \, \int
  d^2 b \, d^2 x \, e^{- i {\bm q} \cdot {\bm x}} \, \nabla^2_x \, N
  ({\bm x}, {\bm b}, y).
\end{align}

It would be very interesting to see if \eq{ktfact} would still hold
when running coupling corrections are included. It would also be
interesting to see whether the relation (\ref{phi}) would survive the
inclusion of running coupling corrections. This would verify the
validity of $k_T$-factorization formula (\ref{ktfact}) and the
universality of the dipole amplitude as the correct degree of freedom
for both the total DIS cross section and for the single gluon
production at small-$x$.

To answer the above questions one has to first understand how to include
running coupling corrections into the gluon production cross section. This is
a formidable task that we are not going to complete here. Instead we will
be content with a first step in that direction by resolving some conceptual
questions and by setting up a formalism for future calculations. Our main task
is to identify the cancellation mechanisms that ensure the IR safety of the
total cross section, as it governs all left over phase space cancellations for
less inclusive observables and allows us to pinpoint the origin of eventual IR
problems.

Below we will analyze the running coupling corrections to the one-gluon
inclusive production cross section. Our strategy is standard: following
Brodsky, Lepage and Mackenzie (BLM) \cite{Brodsky:1983gc} we will begin by
dressing all the relevant gluon lines and vertices by quark bubble chains,
resumming powers of $\as \, N_f$. We will then replace
\begin{align}\label{repl}
N_f \rightarrow - 6 \, \pi \, \beta_2
\end{align} 
to complete it to the full one-loop QCD beta-function
\begin{align}\label{beta}
\beta_2 = \frac{11 N_c - 2 N_f}{12 \, \pi}.  
\end{align} 
Resumming all factors of $\am \, \beta_2$ should replace all factors
of the bare coupling constant $\am$ in the original lowest order
expression (\ref{ktfact}) by the physical couplings.\footnote{We refer
  the reader to
  ~\cite{Mueller:1984vh,Parisi:1978bj,David:1983gz,Zakharov:1992bx}
  and~\cite{Beneke:1998ui,Beneke:2000kc,Beneke:1994qe} for a
  discussion of the context and validity the BLM resummation procedure
  or its refinements, for example via the dispersive method. The first
  issue here is whether the quark bubbles faithfully trace all running
  coupling corrections and thus justify~\eqref{repl}: The answer given
  in the references is that this is true in the ``single dressed gluon
  approximation,'' i.e. if the process only involves one gluon line
  that needs to be dressed. The second issue is the all orders
  resummation involved in renormalizing the coupling: This is deeply
  connected with the large order behavior of perturbation theory and
  the notion of renormalons~\cite{Lautrup:1977hs,'tHooft:1977am}.
  Despite the natural quantitative uncertainties inherent in trying to
  answer the question of what is the scale in the coupling, the BLM
  method should at least yield a qualitatively correct answer and most
  importantly, be able to tell perturbative from non-perturbative
  contributions. If one takes the most conservative and pessimistic
  view, one can always reexpand the resummed coupling back to its
  lowest scale carrying order and estimate the uncertainty induced by
  varying the renormalization scale $\mu$ ``within reasonable
  bounds.'' Our expressions must match these lowest order corrections
  as exemplified for the running coupling corrections to the
  non-linear small-$x$ evolution in the calculation of the gluonic
  component of the NLO BFKL, BK, and JIMWLK kernels done in
  \cite{Balitsky:2008zz}. In \cite{Balitsky:2008zz} the result of a
  full NLO calculation confirmed the scale-setting for the running
  coupling which had been done in
  \cite{Balitsky:2006wa,Kovchegov:2006vj} using the BLM prescription.}

The paper is structured as follows. In Sect. \ref{total} we will set
up the formalism for including quark bubbles in the final state (the
state after the projectile's interaction with the target). We will
then analyze the running coupling corrections to gluon production in
Sect. \ref{glue_mult}.

The main conceptual problem in including running coupling corrections
for the gluon production cross section is with insertion of quark
bubbles onto the outgoing gluon line (see \fig{fig:finalb} below).
Indeed, if one uses dimensional regularization and the quarks are
massless all such bubble corrections are zero, since the on-mass-shell
gluon provides no invariant momentum scale. This zero can be thought
about as a ``cancellation'' of the infrared and ultraviolet
divergences in the bubbles. The zero result is unsatisfactory, since
it would leave the gluon production cross section proportional to a
factor $\am$, the bare coupling, and thus {\em unrenormalized}.

To resolve this problem we notice that collinear divergences,
resulting from gluon splitting into two massless quarks (or gluons)
come in at the same order in perturbation theory as running coupling
corrections on the outgoing gluon line. To keep track of such
divergences we regularize them with the infrared cutoff
$\Lambda_\text{coll}$. With the regulator inserted, we show in Sect.
\ref{glue_mult} that the bubbles on the outgoing gluon line generate a
factor of running coupling $\as (\Lambda_\text{coll}^2)$ with the
infrared (IR) cutoff in the argument. Therefore, the gluon production
cross section has to come in with a factor of the coupling running
with some arbitrary IR cutoff in the argument. Indeed the cutoff has a
meaning of the resolution scale for gluons: once such a cutoff is
introduced, collinear $q\bar q$ pairs (and collinear gluon pairs) are
also identified as gluons, generating extra diagrams contributing
running coupling corrections (see \fig{fig:qqbar_ampl}), which in the
end renormalize gluon production cross section.  We conclude that the
gluon production cross section depends on the IR cutoff/resolution
scale and is thus not infrared-safe, in agreement with
conventional wisdom.

Our argument shows that the result of the full calculation of the running
coupling corrections would necessarily generate a power of $\as
(\Lambda_\text{coll}^2)$. We expect that, again in line with conventional
wisdom, such non-perturbative factors of the coupling could be absorbed into
fragmentation functions if hadron production is the goal of a calculation at
hand. The separation of the diagrams into the perturbative cross sections and
the fragmentation functions could be done using the conventional approach (see
e.g.  \cite{Berger:1996vy,Berger:1995fm} and references therein). It would
result in a factorization scale dependence for both the perturbative cross
section and the fragmentation function.

Indeed one might worry that when \eq{ktfact} is applied to $pA$
collisions, the unintegrated gluon distribution in the projectile
proton $\phi_p$ would depend on some non-perturbative scale $\Lambda$
characterizing the proton, thus making the gluon production cross
section in \eq{ktfact} non-perturbative even without the inclusion of
running coupling corrections. This is a legitimate concern, though,
for instance, it does not apply to the case of DIS, where the scale
characterizing the projectile is the photon's virtuality $Q^2 \gg
\Lambda_\text{QCD}^2$, which makes both $\phi_p$ and the cross section
in \eq{ktfact} perturbative before the inclusion of running coupling
corrections. However, even in the case of DIS, running coupling
corrections would still lead to the collinear singularities that we
discuss below, making the gluon production cross section not
infrared-safe. The problem of applying \eq{ktfact} to $pA$ collisions
is expected to be remedied in nucleus-nucleus ($AA$) collisions or in
$pA$ collisions at very high energies, where the IR-safety of the
gluon production cross section at fixed coupling is expected to be
imposed by the saturation effects in both the target and the
projectile (see
\cite{Kovner:1995ja,Kovner:1995ts,Gyulassy:1997vt,Kovchegov:1997ke}).
While we can not prove that here, we also expect the collinear
singularities discussed below to arise if one tries to calculate the
running coupling corrections to the gluon production in $AA$
collisions.

We continue our discussion by examining an observable built from quark
and gluon production cross sections which is infrared-safe in Sect.
\ref{energy}. The goal is to show explicitly how the infrared
contributions cancel to render the quantity infrared-safe. The
observable is the energy density of the matter produced in a
collision. In \cite{Kovchegov:2005ss} the energy density of the medium
produced in a heavy ion collision at the space-time point described by
the proper time $\tau$, space-time rapidity $\eta$ and transverse
coordinate ${\bm b} = (b_1, b_2)$ was shown to be given by the
following formula:
\begin{align}\label{e}
  \epsilon (\tau, \eta, {\bm b}) \, = \, \frac{1}{\tau} \, \int d^2 p
  \, p_T \, \left\{ \frac{d N^G}{d^2 p \, d \eta \, d^2 b} + \frac{d
      N^q}{d^2 p \, d \eta \, d^2 b} + \frac{d N^{\bar q}}{d^2 p \, d
      \eta \, d^2 b} \right\}.
\end{align}
Here $\frac{d N^G}{d^2 p \, d y \, d^2 b}$ is the number of gluons
produced with transverse momentum $\bm p$, momentum-space rapidity $y$
and at transverse coordinate $\bm b$. The quark and anti-quark
production contributions $\frac{d N^q}{d^2 p \, d \eta \, d^2 b}$ and
$\frac{d N^{\bar q}}{d^2 p \, d \eta \, d^2 b}$ are added to the gluon
one on the right hand side of \eq{e}. Note that the momentum space
rapidity $y$ gets replaced by the space-time rapidity $\eta$ in
\eq{e}.  The above formula (\ref{e}) is valid for sufficiently late
proper times, $\tau \gg 1/\langle p_T \rangle$ with $\langle p_T
\rangle$ the typical transverse momentum of the produced quarks and
gluons. The corrections to it are suppressed by extra powers of $\tau$
\cite{Kovchegov:2005ss}. \eq{e}, along with our discussion below, is
applicable to the particles produced at central rapidity, i.e.,
particles with small-$x$ both in the target and the projectile wave
functions.  \eq{e} is valid not only for heavy ion collisions, but for
$pp$, $pA$ and DIS collisions as well.

In Sect. \ref{energy} we show that the energy density (\ref{e}) is indeed
independent of the IR cutoff $\Lambda_\text{coll}$ by showing that terms in
the gluon (and quark) multiplicity which depend of this cutoff and which are
singular in the $\Lambda_\text{coll} \rightarrow 0$ limit actually vanish when
inserted in \eq{e} using the abstract structure of the equation. In Sect.
\ref{toy_sect} we demonstrate how these terms vanish in an explicit toy model:
we show that collinearly divergent terms have such a $p_T$-dependence, that,
when inserted in \eq{e}, they integrate out to zero. Thus energy density is an
infrared-finite quantity.

We also argue that the running coupling corrections are included in
the expression for the energy density (\ref{e}) in a manner similar to
the total cross section calculation
\cite{Balitsky:2006wa,Kovchegov:2006vj,Kovchegov:2006wf,Albacete:2007yr},
since both gluons and quarks are allowed in the final state for either
the total cross section or for the energy density.

As an aside we discuss the renormalization of the interaction of the
projectile wave function with the target in Appendix \ref{p-np} using
a quark-antiquark dipole as the projectile. We find how the running
coupling corrections enter the Glauber-Mueller formula (\ref{N0})
\cite{Mueller:1989st}. We obtain that one of the two couplings in the
exponent of \eq{N0} runs with the inverse transverse size of the
dipole, while the other coupling runs with some non-perturbative scale
characterizing nucleons in the nucleus, as shown in \eq{N0rc}.

We conclude in Sect. \ref{out} by discussing our main results and
potential future developments.


\section{Total Cross-Section: Cancellation of Fermion Bubbles in the Final State}
\label{total}

We begin by analyzing the running coupling corrections to the total
cross section arising in the final state. Our discussion will apply to
both the cross sections calculated in the quasi-classical
approximation of McLerran-Venugopalan (MV) model
\cite{McLerran:1993ni,McLerran:1993ka,McLerran:1994vd} and to the
cross sections where the corrections due to non-linear small-$x$
evolution equations \cite{Balitsky:1996ub,
  Balitsky:1997mk,Balitsky:1998ya,Kovchegov:1999yj,
  Kovchegov:1999ua,Jalilian-Marian:1997jx, Jalilian-Marian:1997gr,
  Jalilian-Marian:1997dw, Jalilian-Marian:1998cb, Kovner:2000pt,
  Weigert:2000gi, Iancu:2000hn,Ferreiro:2001qy} are included. Just
like in the calculation of the total dipole-nucleus cross section, the
corrections due to small-$x$ evolution come in through a sequence of
longitudinally soft gluon emissions, as only gluonic corrections
generate leading logarithms of Bjorken $x$ (i.e. powers of $\as \, \ln
1/x$) \cite{Kuraev:1977fs,Bal-Lip}. Our discussion below can equally
be applied to the quasi-classical gluon production in the MV model, or
to the gluon production enhanced by small-$x$ evolution corrections,
which appears as a quasi-classical gluon emission in one rung of the
small-$x$ evolution (see e.g.
\cite{Kovchegov:2001sc,Jalilian-Marian:2005jf}). Common to both cases
is an eikonal emission vertex for the gluons under consideration.


\begin{figure}[htb]
  \centering
  \includegraphics[height=4.8cm]{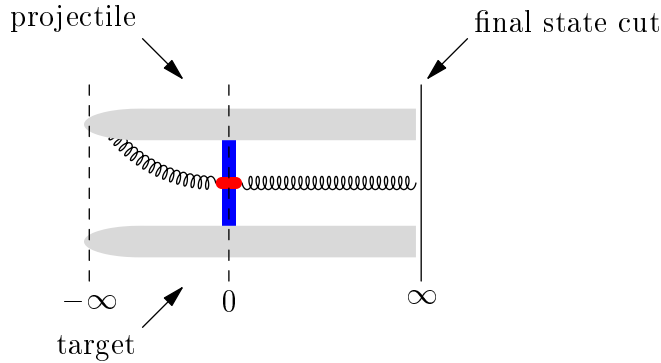}
  \caption{The time-ordered picture of projectile-target scattering amplitude
    introducing notations to be used below. All multiple rescatterings
    are denoted by a broad (blue) vertical line. The solid narrow
    vertical line denotes the final state cut. (Color on-line.)}
  \label{fig:notations}
\end{figure}

In
\cite{Kovchegov:2006vj,Balitsky:2006wa,Kovchegov:2006wf,Gardi:2006rp}
the running coupling corrections to the non-linear small-$x$ evolution
equations \cite{Kovchegov:1999yj, Kovchegov:1999ua, Balitsky:1996ub,
  Balitsky:1997mk, Balitsky:1998ya,Jalilian-Marian:1997xn,
  Jalilian-Marian:1997jx, Jalilian-Marian:1997gr,
  Jalilian-Marian:1997dw, Jalilian-Marian:1998cb, Kovner:2000pt,
  Weigert:2000gi, Iancu:2000hn,Ferreiro:2001qy} were calculated for
the first time. The BK and JIMWLK evolution equations were written for
the total cross section of a projectile (dipole) scattering on a
target: therefore the optical theorem is applicable and allows to
neglect all the final state interactions. This was used
in~\cite{Kovchegov:2006vj,Kovchegov:2006wf} to specifically discard
the final state running coupling corrections.
In~\cite{Balitsky:2006wa} one-loop final state corrections were
studied at order $\as \, N_f$ and their cancellation was shown
explicitly at that level.  While we indeed expect also the all orders
resummed final state fermion bubble insertions to cancel due to the
optical theorem, we expect the explicit mechanism of this cancellation
to be far from straightforward. Thus we set out to explore how the
cancellation works explicitly.

Before we begin, let us introduce the notations. We will consider a
high energy proton-nucleus or DIS collision and will work in the
center of mass frame. In this frame the incoming proton/$q\bar q$
dipole (henceforth referred to as the projectile) comes in from the
light cone time of $x_+ = - \infty$ moving along the $x_+$-axis. The
wave function of the projectile may contain extra gluonic
fluctuations. The projectile scatters on the target instantaneously at
the light cone time $x_+ =0$. Similar to
\cite{Kovchegov:2001sc,JalilianMarian:2004da,Kovchegov:2006qn} we will
denote the instantaneous interaction with the target by a broad (blue)
vertical line, as shown in \fig{fig:notations}.  The broad (blue) line
comprises all possible multiple one- and two-gluon exchanges between
the nucleons in the target and the projectile (similar to
\fig{fig:mugla} in the Appendix \ref{p-np}).  These interactions of
the target field with projectile constituents or the emitted gluon
shown explicitly sum into path ordered exponentials along the
projectile light cone direction. They are indicated for the emitted
gluon (or any explicitly shown interacting quark in what is to follow)
in form of an oval (red) mark.  After scattering on the target the
system evolves further towards the final state at light cone time $x_+
= + \infty$. This final state we will denote by a vertical solid line
as shown in \fig{fig:notations} and refer to it as the final state
cut, as appropriate when talking about diagrammatic contributions to
probabilities instead of amplitudes.

\begin{figure}[htb]
  \centering
  \includegraphics[width=\textwidth]{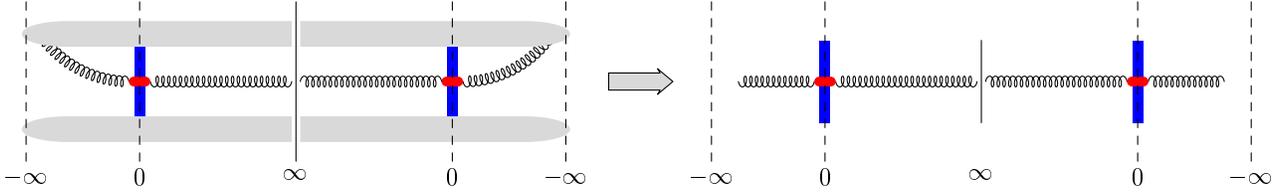}
  \caption{On the left: the scattering amplitude squared for the diagram from 
    \fig{fig:notations}. On the right: an abbreviated notation for the
    same amplitude squared.}
  \label{fig:abbrev}
\end{figure}

The amplitude squared, using a similar notation, is shown in the center
of \fig{fig:abbrev}.
There the light cone time goes from $x_+ = - \infty$ to $x_+ = +
\infty$ both in the amplitude and in its complex conjugate.  In the
case of the projectile dipole one has to sum over all possible
emissions of the gluon by the quark and the anti-quark, both in the
amplitude and in the complex conjugate amplitude. In case of the
proton projectile the emissions off all of the valence quarks have to
be resummed. To save space we will introduce an abbreviated notation,
shown on the right hand side of \fig{fig:abbrev}. The disconnected
gluon line implies sum over emissions of the gluon by all quarks in
the projectile. The broad (blue) line indicates all multiple
rescatterings in the amplitude. This notation also makes it easier to
discuss our results in relation to any type of (small) projectile.

In the notation of the graph on the right of \fig{fig:abbrev} the optical
theorem can be formulated as follows: to calculate the total scattering cross
section the dynamics between $x_+ =0$ and $x_+ = + \infty$ can be ignored. For
instance, if we wanted to find contribution of extra softer gluon emissions to
the total cross section, only emissions at light cone time $-\infty < x_+ <0$
should be included both in the amplitude and in the complex conjugate
amplitude. This indeed has been verified in the literature
\cite{Chen:1995pa,Kovchegov:2001sc}. The same property applies to running
coupling corrections. The quark bubbles dressing the gluon line at times $0 <
x_+ < +\infty$ do not contribute to the total cross section according to the
optical theorem. Such corrections were indeed not included in the calculations
of \cite{Kovchegov:2006vj,Balitsky:2006wa,Kovchegov:2006wf,Gardi:2006rp}.
However, these corrections should be included if one is interested in gluon
production cross section. To learn how to best organize a calculation that
includes them, we will first see how they cancel in the total cross section.

To set the stage, let us briefly recall how the cancellations for the cross
section mandated by the optical theorem work at leading order. The diagrams to
consider for the amplitude include all contributions from order $g^0$ through
$g^2$:
\begin{align}
  \label{eq:lo-amp-sum}
  \parbox{2cm}{\includegraphics[width=2cm]{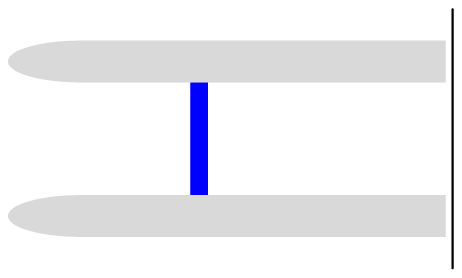}}
  +\parbox{2cm}{\includegraphics[width=2cm]{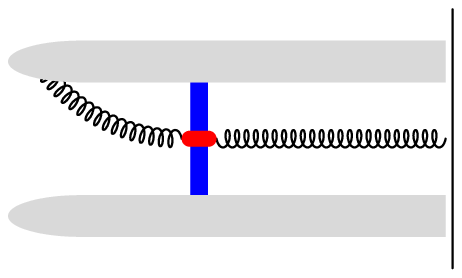}}
  +\parbox{2cm}{\includegraphics[width=2cm]{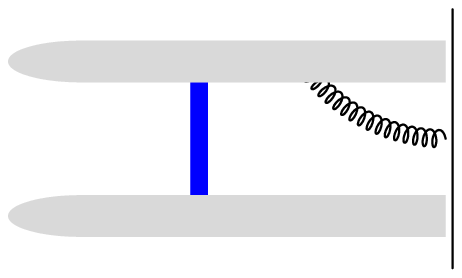}}
  +\parbox{2cm}{\includegraphics[width=2cm]{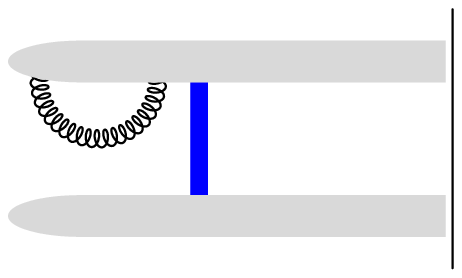}}
 +\parbox{2cm}{\includegraphics[width=2cm]{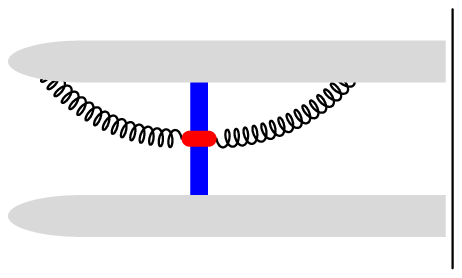}}
 +\parbox{2cm}{\includegraphics[width=2cm]{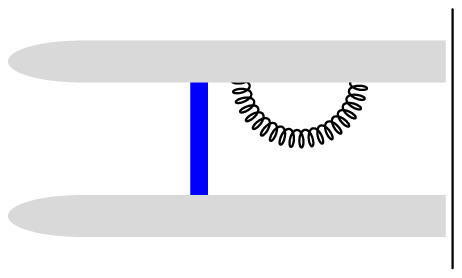}}
\end{align}
The optical theorem guarantees that at order $\alpha_s$, the only diagrams
contributing to the absolute value squared of this amplitude are
\begin{align}
  \label{eq:leading-contrib}
  \parbox{4cm}{\includegraphics[height=1.18cm]{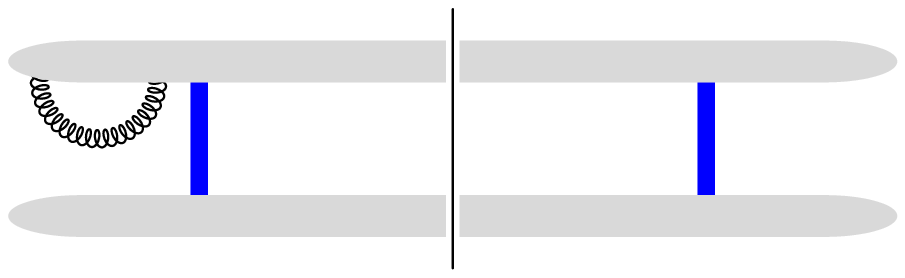}}
  +\parbox{4cm}{\includegraphics[height=1.18cm]{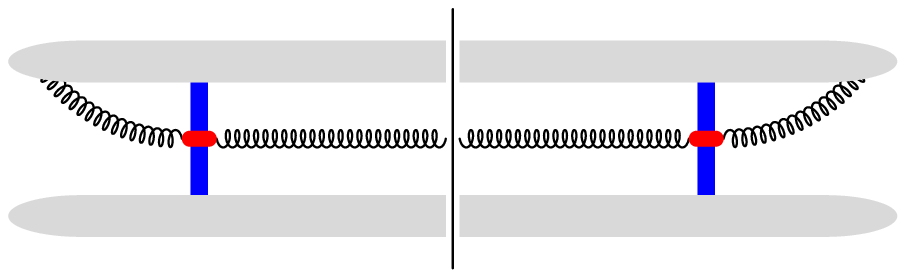}}
  +\parbox{4cm}{\includegraphics[height=1.18cm]{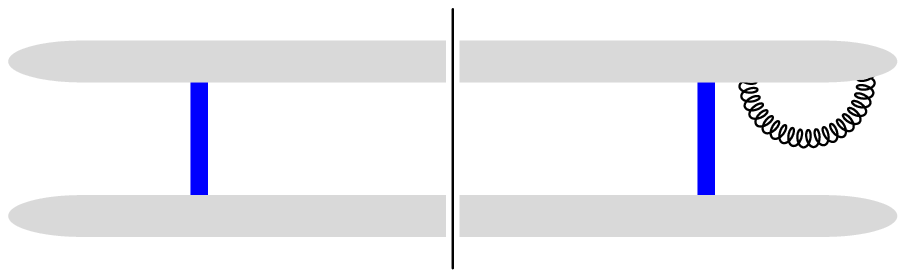}}
\ .
\end{align}
This quite drastic reduction in the number of diagrams one needs to calculate
to obtain the total cross section is due to the following set of
\emph{separate} cancellations:
\begin{subequations}
    \label{eq:leading-order-cancellations}
  \begin{align}
     \parbox{4cm}{\includegraphics[height=1.18cm]{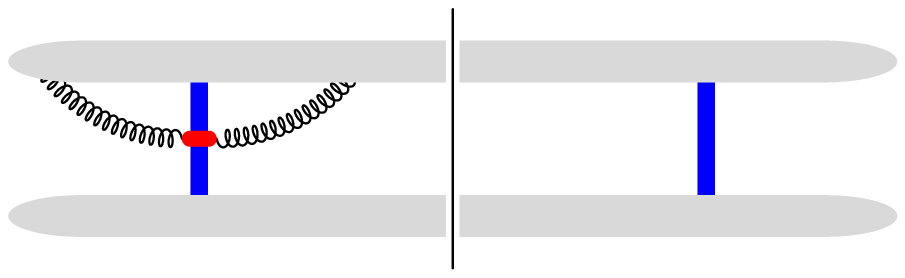}}
  +\parbox{4cm}{\includegraphics[height=1.18cm]{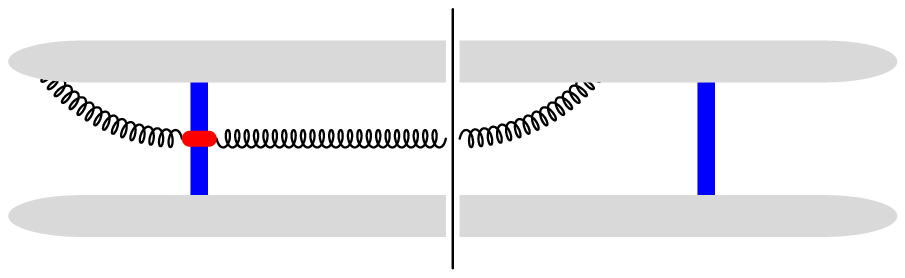}}
\hspace{1.8mm}\phantom{+
\parbox{4cm}{\includegraphics[height=1.18cm]{oc_lc_2}}}
 = & \ 0
\\
\parbox{4cm}{\includegraphics[height=1.18cm]{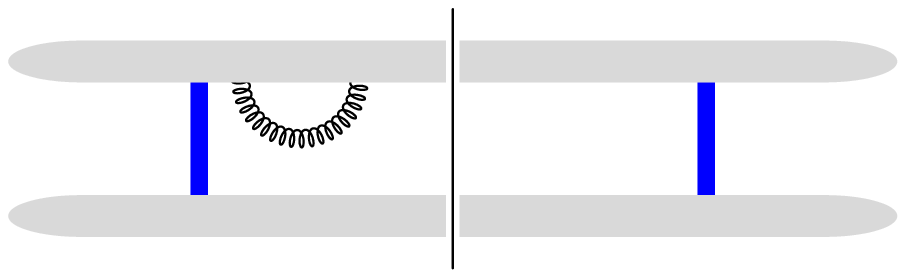}}
  +\parbox{4cm}{\includegraphics[height=1.18cm]{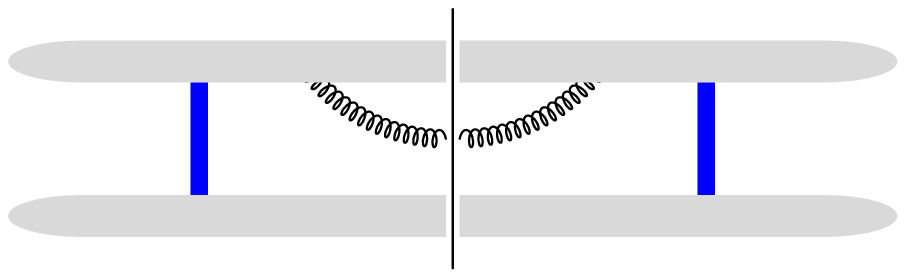}}
  +\parbox{4cm}{\includegraphics[height=1.18cm]{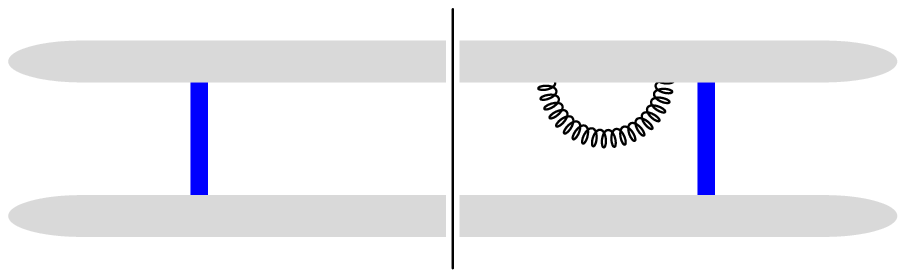}} = & \ 0
\\
\phantom{\parbox{4cm}{\includegraphics[height=1.18cm]{oc_mc_1}}
  +}\parbox{4cm}{\includegraphics[height=1.18cm]{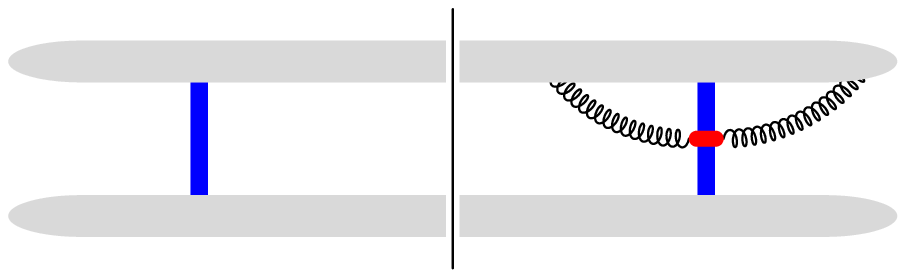}}
  +\parbox{4cm}{\includegraphics[height=1.18cm]{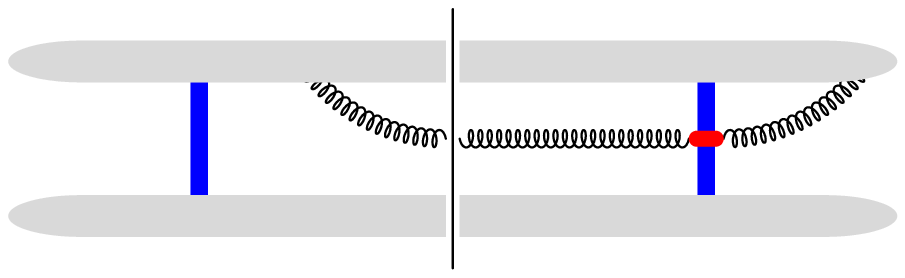}} = & \ 0
\ .
  \end{align}
\end{subequations}
That the cancellation pattern is separated out into three groups is dictated
by the interaction of the produced gluon with the target as marked by the
(red) dot on the interaction area: only diagrams with identical interaction
patterns can possibly cancel against each other. This will be the first
guiding principle to identify potential cancellations below. The actual
cancellations require further analysis of the integrals associated with the
diagrams. If one restricts the sum over the final state phase space, the
cancellations are no longer complete. Still, partial cancellations from the
phase space regions that have not been excluded from the final state sum
remain in effect and only become manifest if one groups the diagrams as shown
in~(\ref{eq:leading-order-cancellations}).

We will not analyze all contributions exhaustively but rather focus on
a number of subsets of the one and two loop fermion bubble insertions
that cancel separately in the total cross section to illustrate and
identify the basic mechanisms at work. This will lay the groundwork to
allow us to precisely pinpoint the origin of IR divergences,
understand how they cancel in the total cross section. It also allows
to understand, for example, the cancellations of UV logarithms of the
form $\ln(s/\mu)$ present in individual diagrams (with $\mu$ the
renormalization scale, a UV quantity) from the final result. All these
relationships are a prerequisite for a full calculation of the running
coupling corrections that remains beyond the scope of the present
paper.

\begin{figure}[htb]
  \centering
  \includegraphics[width=8cm]{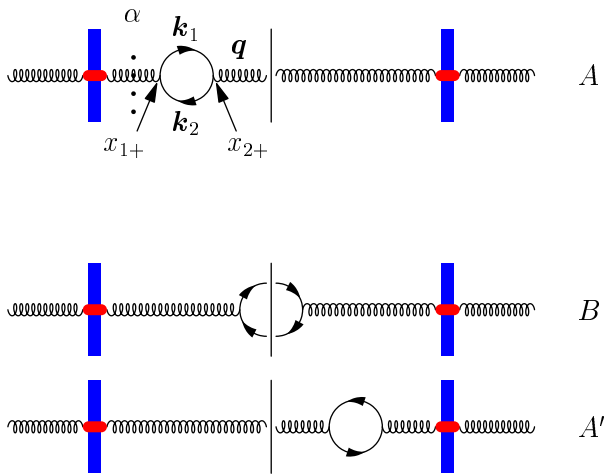}
  \caption{One-loop quark bubble final-state corrections to the contribution
    of the diagram in \fig{fig:abbrev} to the total cross section. Here the
    quark bubble does not interact with the target. The dotted vertical line
    denotes an intermediate state labeled by $\alpha$.}
  \label{fig:ABA}
\end{figure}

\begin{figure}[htb]
  \centering
   \includegraphics[width=8cm]{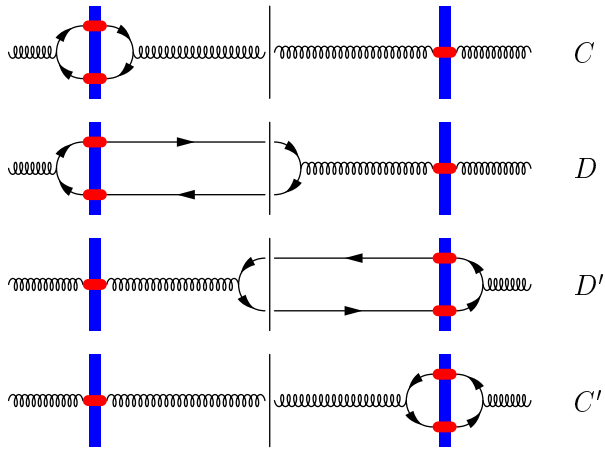}
  \caption{One-loop quark bubble final-state corrections to 
    the contribution of the diagram in \fig{fig:abbrev} to the total
    cross section: here the quark loop interacts with the target.}
  \label{fig:CDDC}
\end{figure}

We begin with a single fermion bubble. The relevant diagrams with
final state interactions (splittings and mergers) are shown in Figs.
\ref{fig:ABA} and \ref{fig:CDDC}.
Due to light-cone time-ordered nature of these graphs, it is more
convenient to use light-cone perturbation theory (LCPT) rules
\cite{Lepage:1980fj,Brodsky:1997de} to calculate them. To understand
the cancellations we will concentrate almost solely on the energy
denominators: apart from complex conjugation on the right hand side of
the cut, the numerators of all diagrams in \fig{fig:ABA} are the same.
At the same time direct application of the rules of LCPT as stated in
\cite{Lepage:1980fj,Brodsky:1997de} appears problematic due to
singularities in some of the energy denominators. One of the singular
denominators corresponds to the intermediate state $\alpha$ shown in
graph A in \fig{fig:ABA}. The light cone energy of the state $\alpha$
is identical to that of the final (outgoing) state, generating a
$1/0$-type singularity coming from the corresponding energy
denominator. This problem has been discussed before in
\cite{Chen:1995pa} for small-$x$ evolution, and our discussion here
will follow the calculations performed in \cite{Chen:1995pa}. To
tackle the problem of singularities one has to calculate the energy
denominators in the diagram in \fig{fig:ABA}A starting from the usual
Feynman perturbation theory.

In going from Feynman perturbation theory to LCPT one performs
integration over the minus components of the internal momenta in the
diagram. A Fourier transform of a denominator of a gluon or quark
propagator becomes
\begin{align}\label{prop}
  \int \frac{d^4 k}{(2 \pi)^4} \, \frac{-i}{k^2 + i \epsilon} \, e^{-
    i k \cdot x} \, = \, \int \frac{d^2 k \, d k_+}{(2 \pi)^3 \, 2
    k_+} \, e^{- i \left( k_+ x_- + \frac{{\bm k}^2 - i \epsilon}{2 \,
        k_+} x_+ - {\bm k} \cdot {\bm x} \right)} \, \left[ - \theta
    (x_+) \, \theta (k_+) + \theta (-x_+) \, \theta (-k_+) \right]
\end{align}
which effectively puts the particle on mass-shell assigning the
light-cone energy (the minus component of the momentum) to be equal to
\begin{align}
  E_k \, = \, \frac{{\bm k}^2}{2 \, k_+}.
\end{align}
Thus a particle propagating over a light cone time interval $x_+$
brings in a factor of
\begin{align}
  e^{- i E_k \, x_+}.
\end{align}
The incoming particle with momentum $(k_+, {\bm k})$ brings in a
factor of
\begin{align}
  e^{- i E_k \, x_+}
\end{align}
while the outgoing particle with momentum $(k_+, {\bm k})$ brings in
\begin{align}
  e^{i E_k \, x_+},
\end{align}
as can be inferred from \eq{prop}. 

With the above rules in mind, the relevant contribution of the diagram
$A$ in \fig{fig:ABA} is
\begin{align}\label{3A1}
  A \, = \, - \int\limits_0^{+\infty} d x_{2+} \,
  \int\limits_0^{x_{2+}} d x_{1+} \, e^{- \delta x_{1+} - \delta
    x_{2+} - i (x_{2+} - x_{1+}) (E_{k_1} + E_{k_2}) - i x_{1+} E_q +
    i x_{2+} E_q },
\end{align}
where the overall minus sign comes from taking into account two powers
of $i$ coming from the quark-gluon vertices. The omitted (numerator)
part of the diagram $A$ is the same for all three graphs in
\fig{fig:ABA}: the gluon in all three cases can only have transverse
polarizations. In \eq{3A1} we have introduced light-cone regulators
$\delta > 0$, which make the integrals convergent at infinite light
cone time. (Indeed, for the initial state emissions the regulators
should come in with the opposite sign in the exponent.)  The notations
used in \eq{3A1} are explained in \fig{fig:ABA}A: the quark and
anti-quark lines carry momenta $(k_{1+}, {\bm k}_1)$ and $(k_{2+},
{\bm k}_2)$ correspondingly, while the gluon carries momentum $(q_{+},
{\bm q}) = (k_{1+}, {\bm k}_1) + (k_{2+}, {\bm k}_2)$. The quark-gluon
vertices are assigned light cone times $x_{1+}$ and $x_{2+}$, as shown
in \fig{fig:ABA}A. Due to theta-functions in \eq{prop} one can see
that only the ordering $0 \le x_{1+} \le x_{2+} < +\infty$ used in
\eq{3A1} is allowed, as $q_+ >0$ for the produced gluon.

Performing the integrations in \eq{3A1} and expanding the result in
the powers of $\delta$ yields
\begin{align}\label{3A}
  A \, = \, - \frac{i}{2 \, \delta} \, \frac{1}{E_q - E_{k_1} -
    E_{k_2}} - \frac{1}{2} \, \frac{1}{(E_q - E_{k_1} - E_{k_2})^2} +
  o(\delta).
\end{align}
Noticing that the diagram $A'$ in \fig{fig:ABA} is equal to a complex
conjugate of the diagram $A$ we obtain
\begin{align}\label{AA}
  A + A' \, = \, A + A^* \, = \, - \frac{1}{(E_q - E_{k_1} -
    E_{k_2})^2} + o(\delta).
\end{align}
($A^*$ denotes a complex conjugate of $A$.)  The dangerous $1/\delta$
singularity is canceled in the sum of the two diagrams.  Calculating
the diagram B in \fig{fig:ABA} in a similar manner (or directly by using LCPT
rules from \cite{Lepage:1980fj,Brodsky:1997de} since there is no
singular energy denominators in this graph) one gets
\begin{align}\label{3B}
  B \, = \, \frac{1}{(E_q - E_{k_1} - E_{k_2})^2} + o(\delta).
\end{align}
Adding Eqs. (\ref{AA}) and (\ref{3B}) and taking the $\delta
\rightarrow 0$ limit yields
\begin{align}\label{AAB}
  A + A' + B = 0. 
\end{align}
We have thus shown the cancellation of the diagrams in \fig{fig:ABA} for the
calculation of the total cross section.

Diagrams C and D (and C' and D') in \fig{fig:CDDC}
are in a separate class of diagrams where the quark loop interacts with the
target. They do not have singular energy denominators. Applying the LCPT rules
\cite{Lepage:1980fj,Brodsky:1997de} one obtains for the relevant parts of
those graphs:
\begin{align}
  C \, = \, \frac{1}{E_{k_1} + E_{k_2} - E_q}
\end{align}
and 
\begin{align}
  D \, = \, \frac{1}{E_q - E_{k_1} - E_{k_2}},
\end{align}
such that 
\begin{align}\label{CD}
  C + D = 0. 
\end{align}
Similarly 
\begin{align}\label{CD'}
  C' + D' = 0. 
\end{align}
This accomplishes the proof of the cancellation of one-fermion bubble
corrections in the final state for the total cross section. We observe
that cancellations happen within classes of diagrams that share the
same type of interactions with the target: diagrams with a gluon
interacting with the target cancel separately from the diagrams with
the quark bubble interacting with the target.

\begin{figure}[htb]
  \centering
   \includegraphics[width=8cm]{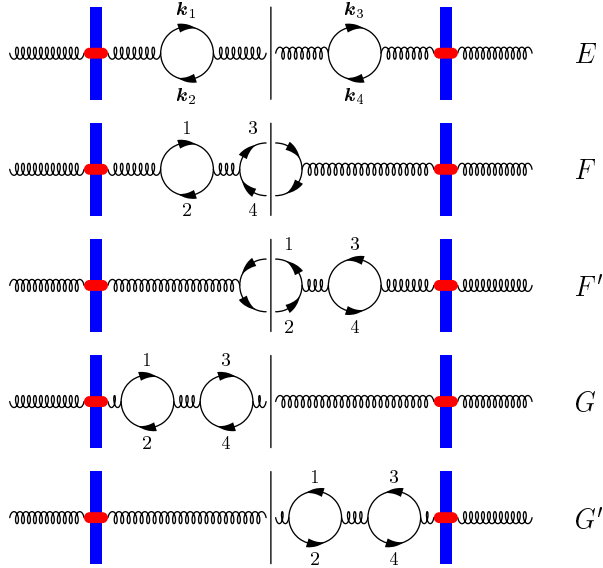}
  \caption{Two-loop quark bubble final-state corrections to 
    the contribution of the diagram in \fig{fig:abbrev} to the total
    cross section. Here the quark bubbles do not interact with the
    target.}
  \label{fig:EFG}
\end{figure}

\begin{figure}[hbt]
  \centering
   \includegraphics[width=8cm]{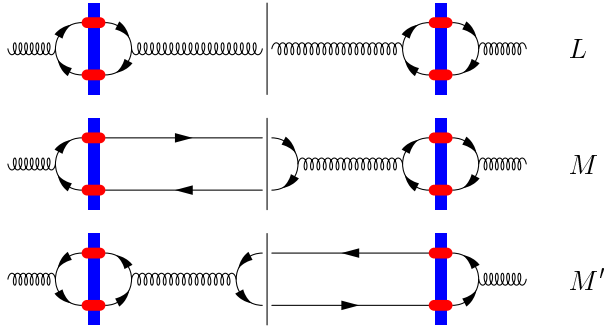}
  \caption{Two-loop quark bubble final-state corrections to 
    the contribution of the diagram in \fig{fig:abbrev} to the total
    cross section: both bubbles interact with the target. }
  \label{fig:LMM}
\end{figure}

To see if the pattern persists at higher orders, let us analyze the diagrams
with two fermion bubbles. The different classes with separate cancellation
patterns are shown in figs.~\ref{fig:EFG},~\ref{fig:HIJK} and~\ref{fig:LMM}.
We begin with diagrams that have only the gluon interacting with the target
shown in~\fig{fig:EFG}.
Their calculation proceeds along the lines outlined above. Note, that
the diagram E in \fig{fig:EFG} is a square of the absolute value of
the diagram A in \fig{fig:ABA}: however, in order to keep all the
finite pieces of diagram E one has to expand the expression for the
diagram A up to order $\delta$ before squaring it (i.e., one order
higher than shown in \eq{3A}), since the $1/\delta$ singularities make
order $\delta$ corrections in A contribute to E.  The final result
yields
\begin{align}
  E \, = & \, \frac{1}{4 \, \delta^2} \, \frac{1}{(E_q - E_{k_1} -
    E_{k_2}) \, (E_q - E_{k_3} - E_{k_4})} - \frac{i}{4 \, \delta} \,
  \frac{E_{k_1} + E_{k_2}- E_{k_3} - E_{k_4}}{(E_q - E_{k_1} -
    E_{k_2})^2 \, (E_q - E_{k_3} - E_{k_4})^2} \notag \\
  & + \frac{1}{4} \left[ - \frac{1}{(E_q - E_{k_1} - E_{k_2}) \, (E_q
      - E_{k_3} - E_{k_4})^3} + \frac{1}{(E_q - E_{k_1} - E_{k_2})^2
      \, (E_q - E_{k_3} - E_{k_4})^2} \right. \notag \\ & \left.  -
    \frac{1}{(E_q - E_{k_1} - E_{k_2})^3 \, (E_q - E_{k_3} - E_{k_4})}
  \right] + o(\delta).
\end{align}

For the other diagrams in \fig{fig:EFG} one gets
\begin{align}
  F \, = \, \frac{1}{(E_q - E_{k_1} - E_{k_2})^3 \, (E_{k_3} +
    E_{k_4}- E_{k_1} - E_{k_2})},
\end{align}
\begin{align}
  F' \, = \, F^* (k_1, k_2 \leftrightarrow k_3, k_4),
\end{align}
\begin{align}
  G \, = & \, - \frac{1}{8 \, \delta^2} \, \frac{1}{(E_q - E_{k_1} -
    E_{k_2}) \, (E_q - E_{k_3} - E_{k_4})} - \frac{i}{8 \, \delta} \,
  \frac{E_{k_1} + E_{k_2} + 3 E_{k_3} + 3 E_{k_4} - 4 E_q}{(E_q -
    E_{k_1} - E_{k_2})^2 \, (E_q - E_{k_3} - E_{k_4})^2} \notag \\
  & + \frac{9}{8} \, \frac{1}{(E_q - E_{k_1} - E_{k_2}) \, (E_q -
    E_{k_3} - E_{k_4})^3} + \frac{3}{8} \, \frac{1}{(E_q - E_{k_1} -
    E_{k_2})^2 \, (E_q - E_{k_3} - E_{k_4})^2} \notag \\ & +
  \frac{9}{8} \, \frac{1}{(E_q - E_{k_1} - E_{k_2})^3 \, (E_q -
    E_{k_3} - E_{k_4})},
\end{align}
\begin{align}
  G' \, = \, G^* (k_1, k_2 \leftrightarrow k_3, k_4).
\end{align}
Adding the contributions up yields
\begin{align}
  E + F + F' + G + G' =0.
\end{align}

Similarly one can demonstrate the cancellation of the diagrams with
the quark bubble interacting with the target on one side of the cut
and the gluon interacting with the target on the other side of the
cut, as shown in \fig{fig:HIJK}:
\begin{figure}[htb]
  \centering
   \includegraphics[width=8cm]{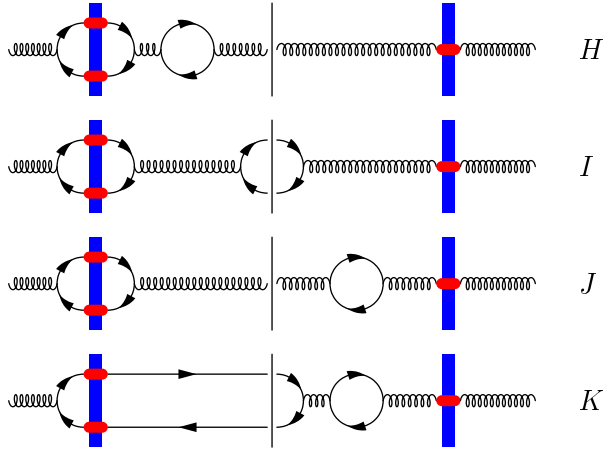}
  \caption{Two-loop quark bubble final-state corrections to 
    the contribution of the diagram in \fig{fig:abbrev} to the total
    cross section: here only one of the bubbles interacts with the
    target. }
  \label{fig:HIJK}
\end{figure}
\begin{align}
  H + I + J + K =0. 
\end{align}
(The same also applies to the mirror reflection of those graphs with
respect to the cut.)

Finally, the diagrams with quark bubbles interacting with the target
on both sides of the cut shown in \fig{fig:LMM} 
also cancel:
\begin{align}
  L + M + M' =0.
\end{align}

We have thus demonstrated cancellation of fermion bubbles in the final
state up to the two-loop level for the total cross section. We
conclude that such cancellations happen at all orders. As was noted
before, due to the optical theorem, no final state interactions enter
the total cross section. Alternatively one could argue that for the
total cross section it is not important what happens to the gluon
after the interaction with the target: it may split into a $q\bar q$
pair, like in \fig{fig:ABA}B, or the $q\bar q$ pair may recombine back
into a gluon, as in \fig{fig:ABA}A. Due to probability conservation,
the contributions of those two events should cancel, as the gluon
remains a single gluon with probability one. This conclusion is true
for any number of bubbles on the gluon line -- all of them should
cancel after the sum over all cuts.

The above cancellations happen only if the momenta of the appropriate
quark and gluon lines are equal in all canceling diagrams\footnote{We
  thank Jianwei Qiu for asking one of us a question which led us to
  the understanding of the subtle effects discussed below.}. For
instance, the momenta ${\bm k}_1$, ${\bm k}_2$, ${\bm q}$ along with
$k_{1+}$, $k_{2+}$, and $q_+$ have to be taken at the same values for
all three diagrams $A$, $B$ and $A'$ in \fig{fig:ABA} for the
cancellation of \eq{AAB} to take place. However, the allowed values of
$k_1$ and $k_2$ are different in the virtual diagrams $A$ and $A'$ and
in the real diagram $B$ in \fig{fig:ABA}. In fact, as one can easily
see, the diagrams $A$ and $A'$ in \fig{fig:ABA} give an ultraviolet
(UV) divergence after the integration over the momentum in the loop.
At the same time the UV divergence in the diagram \ref{fig:ABA}$B$ is
cut off by the center-of-mass energy $s$ in the scattering problem.
Therefore the cancellation in \fig{fig:ABA} is not complete: in fact,
instead of the zero on the right-hand-side of \eq{AAB}, one should get
a contribution proportional to $\ln (s/\mu)$ with $\mu$ the
renormalization scale (a UV cutoff). One would get a paradoxical
result: quark loop of \fig{fig:ABA} would bring in a leading logarithm
of the center-of-mass energy $s$ (i.e., a logarithm which comes in
only with one power of the coupling in front). This is indeed
impossible, since it is well known that the leading logarithms of the
center-of-mass energy $s$ are given by the gluonic contribution only
\cite{Kuraev:1977fs,Bal-Lip}.
\begin{figure}[htb]
  \centering
  \includegraphics[width=8cm]{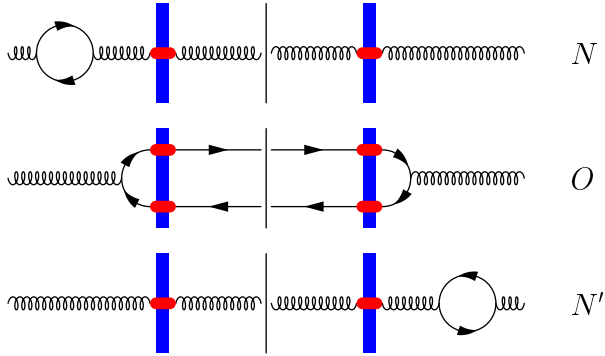}
  \caption{One-loop quark bubble initial-state corrections to 
    the contribution of the diagram in \fig{fig:abbrev} to the total
    cross section.}
  \label{fig:NON}
\end{figure}

The resolution of this apparent paradox is in the fact that, while the
particular set of diagrams in \fig{fig:ABA} does generate an additional
leading logarithm of the center-of-mass energy $s$, such logarithms get
canceled if we sum all the relevant diagrams, a possibility that has been
encountered already in~\cite{Babansky:2002my}. Consider diagrams in Figs.
\ref{fig:CDDC} and \ref{fig:NON} which exhibit the same problem. For instance,
diagram $C$ in \fig{fig:CDDC} is UV divergent, while diagram $D$ in
\fig{fig:CDDC} is cut off by $s$ in the UV. Since the diagram in
\fig{fig:CDDC}$D$ comes in with the overall minus sign compared to the diagram
in \fig{fig:ABA}$B$, as one of the quark-gluon vertices is moved to early
time, we conclude that diagrams $C$ and $D$ in \fig{fig:CDDC} give a
contribution proportional to $- \ln (s/\mu)$.  In the deep UV limit that we
are studying the interactions of the quark bubble with the target are the same
as the interactions of the gluon with the target: in the UV limit the quark
and the anti-quark in the bubble would be very close to each other in
transverse plane. Hence the remaining factors in the diagrams $C$ and $D$ in
\fig{fig:CDDC} are the same as those in the diagrams $A$, $B$ and $A'$ in
\fig{fig:ABA}. Therefore the logarithms $\ln (s/\mu)$ cancel between diagrams
\ref{fig:ABA} $A$, $B$, $A'$ and diagrams \ref{fig:CDDC} $C$, $D$. Similarly
one can show that the logarithms $\ln (s/\mu)$ cancel between diagrams
\ref{fig:NON} $N$, $O$, $N'$ and diagrams \ref{fig:CDDC} $C'$, $D'$. In the
end, all the extra logarithms of $s$ cancel, just as in the example
of~\cite{Babansky:2002my} for leading-order evolution.

In fact, the above observation applies to the leading-order small-$x$
evolution leading to the JIMWLK and BK equations \cite{Balitsky:1996ub,
  Balitsky:1997mk,Balitsky:1998ya,Kovchegov:1999yj,
  Kovchegov:1999ua,Jalilian-Marian:1997jx, Jalilian-Marian:1997gr,
  Jalilian-Marian:1997dw, Jalilian-Marian:1998cb, Kovner:2000pt,
  Weigert:2000gi, Iancu:2000hn,Ferreiro:2001qy}. There, in the forward
amplitude, the diagrams with the gluon interacting with the target have a UV
divergence, which should be regulated by the center-of-mass energy $s$. The
diagrams with the virtual gluon have a divergence regulated simply by a
cutoff. In the full evolution equation the divergences cancel between the real
and virtual terms. To really see this, one has to gather all diagrams that
become similar to the above in the UV to argue that logarithms of $s$ cancel
as well. One might be tempted to argue that the diagrams like the one shown
in~\fig{fig:ABA} do not contribute to the forward amplitude. While this
is true for generic momenta in the bulk of phase space, it is in
fact {\em not} the case for the deep UV limit, where the transverse momenta of
the quarks in the loop are large, making the light-cone lifetime of the quark
bubble very short, shorter than the time interval between the multiple
rescatterings in the target.  Therefore such diagrams in this UV limit have to
be included in the forward amplitude. However, due to the cancellation
patterns observed above we see that all such deep-UV contributions cancel
exactly (according to Eqs.  (\ref{AAB}), (\ref{CD}), (\ref{CD'}), etc.), with
the net effect that the {\em sums} of diagrams in Figs.  \ref{fig:ABA},
\ref{fig:CDDC}, etc., do not contribute to the forward amplitude, despite the
fact that the {\em individual} diagrams would do so. In reality one has
to be aware of this subtlety and remember that the diagrams under
consideration do not contribute only because of an intricate cancellation.

We will now apply the machinery developed above to the calculation of
running coupling corrections to the inclusive gluon production.


\section{Inclusive Production: Summing Fermion Bubbles on the Outgoing Gluon Line}
\label{glue_mult}

\begin{figure}[htb]
  \centering
  \includegraphics[height=1.4cm]{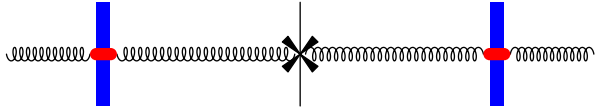}
  \caption{Gluon production diagram for projectile-nucleus scattering using 
    the notation introduced above.}
  \label{fig:bare}
\end{figure}

Now let us try to include running coupling corrections to the
inclusive gluon production in the scattering of a small projectile on
a large target. A diagram contributing to the gluon production cross
section is shown in \fig{fig:bare} 
in the notation introduced in the previous Section. The cross on the
gluon line indicated that we tag on that gluon, keeping its momentum
fixed.

To include the running coupling correction we have to again ``dress''
all gluon lines and vertices with quark bubbles. To begin with let us
point out that $t$-channel gluon exchanges with the target get
renormalized separately from the rest of the diagram: each $t$-channel
gluon with transverse momentum $\bm l$ comes in with a factor of the
bare coupling constant $\am$ and, after being dressed with fermion
bubbles, it acquires the standard geometric series which can be
absorbed in the denominator leading to the physical running coupling
constant
\begin{align}
  \as ({\bm l}^2) \, = \, \frac{\am}{1 + \am \, \beta_2 \, \ln
    \frac{{\bm l}^2}{\mu^2}}.
\end{align}

The details of this calculation are shown in Appendix \ref{p-np},
where we show how running coupling corrections enter into multiple
gluon exchanges. There we consider dipole-nucleus scattering in the
quasi-classical approximation. The results of the Appendix \ref{p-np},
while directly applicable to DIS, can also be applied to gluon-target
interactions shown in \fig{fig:bare}. Using the crossing symmetry one
can show that the gluon-target interaction on both sides of the cut in
\fig{fig:bare} is equivalent to a gluon dipole--target interaction
\cite{Kovchegov:2001sc,Jalilian-Marian:2005jf} with the gluon dipole
made out of the gluon lines on both sides of the cut. At large $N_c$
the gluon dipole is equivalent to two quark dipoles considered in
Appendix \ref{p-np}. We expect the structure of the running coupling
corrections found in Appendix \ref{p-np} for a quark dipole to be
preserved for the gluon dipole beyond large $N_c$ limit.

It is interesting to note that the vertices where the $t$-channel
gluon connects to the nucleons in the target come with a
non-perturbative coupling, whose scale is determined by the typical
momentum scale in each nucleon, i.e., by the scale of order of
$\Lambda_{\text{QCD}}$.  The vertices where the $t$-channel gluons
couple to the projectile or to a gluon in the projectile's wave
function come with the coupling at the scale of the gluon's transverse
momentum and are, therefore, perturbative.

The important message for the discussion below is that the multiple
rescatterings renormalize separately from the rest of the diagram. The
only remaining factor of bare coupling comes from the gluon emission
vertex. In \fig{fig:bare} the gluon is emitted before the interaction on both
sides of the cut.  Since summation over emissions off of all quarks is
implied in \fig{fig:bare}, the emission vertices are not shown explicitly. The
two vertices give a factor of $\am$: our goal is to see how it
transforms into a physical running coupling constant.

\begin{figure}[htb]
  \centering
  \includegraphics[height=1.4cm]{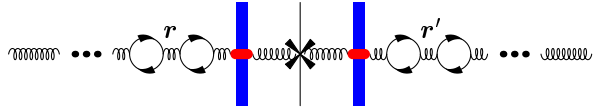}
  \caption{Initial-state running coupling corrections to gluon 
    production cross section. $r$ and $r'$ label the momenta of the
    gluon line on the left and on the right of the cut.}
  \label{fig:initialb}
\end{figure}

The gluon propagators at the light cone time before the interaction,
$x_+ < 0$, can be dressed by quark bubbles both in the amplitude and
in the complex conjugate amplitude, as shown in \fig{fig:initialb}.
Both for Feynman diagrams and in LCPT such quark bubble chains would
give a geometric series each. The two chains of bubbles would
contribute a factor
\begin{align}
  \frac{1}{\left[ 1 + \am \, \beta_2 \, \ln \frac{{\bm r}^2}{\mu^2}
    \right] \, \left[ 1 + \am \, \beta_2 \, \ln \frac{{\bm
          r'}^2}{\mu^2} \right]}
\end{align}
where $\bm r$ and $\bm r'$ are the transverse momenta of the gluon
before interaction with the target in the amplitude and in the complex
conjugate amplitude, as shown in \fig{fig:initialb}.

At the same time the vertices of the gluon interaction with the target
could also have additive quark bubble (vertex) corrections. Adding the
contributions of such bubbles gives a factor
\begin{align}
  \left[ 1 + \am \, \beta_2 \, \ln \frac{{\bm Q}^2}{\mu^2}
    \right] \, \left[ 1 + \am \, \beta_2 \, \ln \frac{{\bm
          Q'}^2}{\mu^2} \right]
\end{align}
where $Q$ and $Q'$ are some momentum scales characterizing the
vertices (which, in the actual calculations are likely to become
combinations of momenta of several gluon lines involved). 

The combination of all bubble corrections in \fig{fig:initialb} and the bare
coupling coming from the emission vertex is
\begin{align}
  \am \, \frac{ \left[ 1 + \am \, \beta_2 \, \ln \frac{{\bm
          Q}^2}{\mu^2} \right] \, \left[ 1 + \am \, \beta_2 \, \ln
      \frac{{\bm Q'}^2}{\mu^2} \right]}{\left[ 1 + \am \, \beta_2 \,
      \ln \frac{{\bm r}^2}{\mu^2} \right] \, \left[ 1 + \am \, \beta_2
      \, \ln \frac{{\bm r'}^2}{\mu^2} \right]} \, = \, \am \,
  \frac{\as (r^2) \, \as (r'^2) }{\as (Q^2) \, \as (Q'^2)}.
\end{align}
One can see that the bubble corrections of \fig{fig:initialb} still fail to turn
the emission vertex coupling constant $\am$ into the physical running
coupling constant. The only remaining bubble corrections which may
complete $\am$ to a physical coupling constant are the bubble
corrections placed on the outgoing gluon line in the final state.

\begin{figure}[htb]
  \centering
    \includegraphics[height=1.4cm]{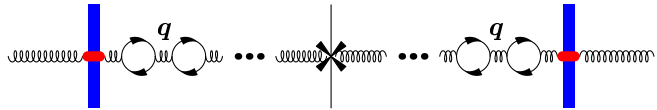}
  \caption{Final-state running coupling corrections to gluon production 
    cross section. The gluon carries momentum $q$. }
  \label{fig:finalb}
\end{figure}

\begin{figure}[h]
 $\displaystyle
    \sum\limits_{
      \begin{minipage}{1cm}
        gluon\\
        cuts
      \end{minipage}
}
\parbox{6.8cm}{\includegraphics[width=6.8cm]{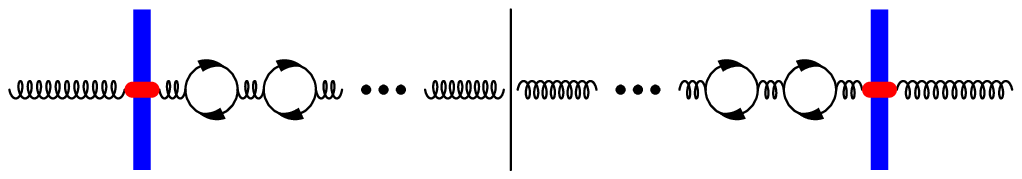}}
= 
    -\!\!\sum\limits_{
      \begin{minipage}{1cm}
        quark\\
        cuts
      \end{minipage}
}
\parbox{6.8cm}{\includegraphics[width=6.8cm]{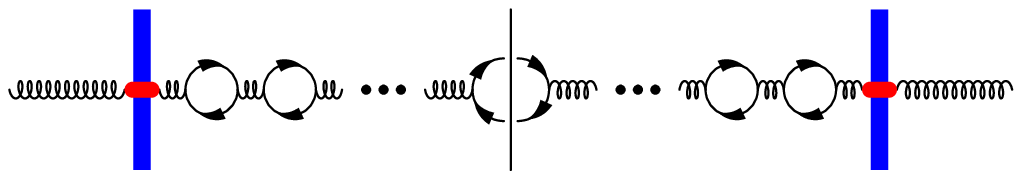}}
$
  \caption{Diagrammatic representation of the consequence of the final-state 
    interaction cancellations observed in Sect. \ref{total}.}
  \label{fig:diag_calc}
\end{figure}

The quark bubble chain corrections on the tagged gluon line are shown
in \fig{fig:finalb}.  To sum them up we will use the result of Sect.
\ref{total} demonstrating that the sum of such corrections with the
ones with the $q\bar q$ pair in the final state cancels. Therefore,
the sum of all the diagrams with the gluon in the final state, like
those shown in \fig{fig:finalb}, is equal to the negative of the sum
of all the diagrams with the $q\bar q$ pair in the final state, as
illustrated in \fig{fig:diag_calc}. The diagrams with the $q\bar q$
pair in the final state (on the right hand side of
\fig{fig:diag_calc}) have the advantage of not generating any singular
energy denominators, like the ones discussed in Sect.  \ref{total}.

\begin{figure}[htb]
  \centering
  \includegraphics[width=7cm]{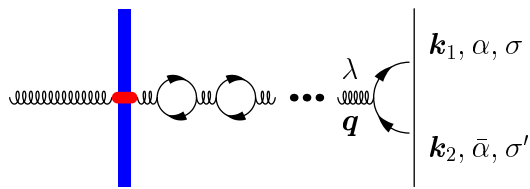}  
  \caption{A diagram with a $q\bar q$ pair in the final state and with 
    running coupling (quark bubble) corrections included on the gluon
    line. The negative square of this diagram gives the
    right-hand-side of \fig{fig:diag_calc}.}
  \label{fig:qqbar_ampl}
\end{figure}

While we can calculate the diagrams on the right hand side of
\fig{fig:diag_calc} using LCPT, it is more straightforward to do so
using the standard Feynman diagram technique. In \fig{fig:qqbar_ampl}
we show the amplitude with the $q\bar q$ pair in the final state and
with the gluon line dressed by quark bubbles. Our discussion below
will be restricted to the case when the gluon is emitted in the
projectile before the interaction, as shown in \fig{fig:qqbar_ampl}.
The conclusions we will obtain will be straightforward to generalize
to the case of gluon emission after the interaction.

Let us consider massless quarks in the bubbles.  In order to regulate
collinear divergences, a study of which is our primary goal here, we
will regulate the integral over the relative momentum of the $q\bar q$
pair in the final state with some infrared cutoff
$\Lambda_{\text{coll}}$.

Completing the factor of $N_f$ coming from the fermion bubbles on the
gluon line to the full beta-function by the replacement $N_f
\rightarrow - 6 \pi \beta_2$ we obtain the following expression for
the amplitude depicted in \fig{fig:qqbar_ampl} (for similar
calculations see
\cite{Kovchegov:2006qn,Kovchegov:2006vj,Kovchegov:1999kx})
\begin{align}\label{ampl1}
  A \, = \, - \sum\limits_{\lambda = \pm 1} {\bm M}^a ({\bm q}, q_+)
  \cdot {\bm \epsilon}^\lambda \, \frac{-i}{q^2 + i \epsilon} \,
  \frac{4 \, \pi \, \am}{1 + \am \, \beta_2 \, \left[ \ln \left( -
        \frac{q^2 + i \epsilon}{\mu_{\overline{\text{MS}}}^2} \right)
      - \frac{5}{3} \right]} \, t^a \, \frac{{\bm \epsilon}^{\lambda
      *} \cdot {\bm k} \, (1 - 2 \alpha + \lambda \sigma) \,
    \delta_{\sigma, \sigma'}}{\alpha \, (1-\alpha) \, q_+}.
\end{align}
The notations used above are explained in \fig{fig:qqbar_ampl}: the outgoing quark
and anti-quark have transverse momenta ${\bm k}_1$ and ${\bm k}_2$,
while the gluon has 4-momentum $q_\mu$. The longitudinal fraction of
the gluon's momentum carried by the quark is
\begin{align}\label{alpha}
  \alpha \, \equiv \, \frac{k_{1+}}{q_+} \, = \, \frac{k_{1+}}{k_{1+}
    + k_{2+}}.
\end{align}
In \eq{ampl1} we have defined
\begin{align}
  {\bm k} \, \equiv \, (1-\alpha) \, {\bm k}_1 - \alpha \, {\bm k}_2.
\end{align}
The quark and anti-quark helicities are labeled $\sigma$ and $\sigma'$
and the gluon polarization vector is $\epsilon_\mu^\lambda = (0,0,{\bm
  \epsilon}^\lambda)$ with ${\bm \epsilon}^\lambda = (1 + i \,
\lambda)/\sqrt{2}$ and $\lambda = \pm 1$. In diagram shown in \fig{fig:qqbar_ampl}
only transverse gluon polarizations contribute. The gluon carries
color $a$ and $t^a$ is the color matrix: summation over repeated
indices is implied.

Remembering that outgoing quark and anti-quark are on mass shell and
massless we write
\begin{align}\label{Ddef}
  q^2 = \frac{{\bm k}^2}{\alpha \, {\bar \alpha}}
\end{align}
with ${\bar \alpha} = 1 - \alpha$.

The two-dimensional vector quantity ${\bm M}^a ({\bm q}, q_+)$ denotes
the part of the amplitude including the gluon emission and its
interaction with the target. The fact that ${\bm M}^a ({\bm q}, q_+)$
depends only on $\bm q$ and $q_+$, and is independent of $q^2$ (or,
equivalently, $\bm k$), is most easily seen if this quantity is
calculated in LCPT. There, all energy denominators in ${\bm M}^a ({\bm
  q}, q_+)$ would not include any $\bm k$-dependence since ${\bm M}^a
({\bm q}, q_+)$ includes only the initial state dynamics and $\bm k$
is the final state momentum. The factor of the coupling constant $g$
due to gluon emission by the projectile is not included in ${\bm M}^a
({\bm q}, q_+)$ and is shown explicitly in \eq{ampl1} as one of the
two factors of $g$ giving $\am$ in the numerator.

As $q^2 = {\bm k}^2 /\alpha \, {\bar \alpha} \ge 0$, and, since later
on we will regulate the collinear divergence at $\bm k=0$ with a
cutoff $\Lambda_{\text{coll}}$, we drop the $i \epsilon$ regulators in
\eq{ampl1} obtaining
\begin{align}\label{ampl2}
  A \, = \, \sum\limits_{\lambda = \pm 1} {\bm M}^a ({\bm q}, q_+)
  \cdot {\bm \epsilon}^\lambda \, \frac{i}{q^2} \, \frac{4 \, \pi \,
    \am}{1 + \am \, \beta_2 \, \left[ \ln \left( -
        \frac{q^2}{\mu_{\overline{\text{MS}}}^2} \right) - \frac{5}{3}
    \right]} \, t^a \, \frac{{\bm \epsilon}^{\lambda *} \cdot {\bm k}
    \, (1 - 2 \alpha + \lambda \sigma) \, \delta_{\sigma, \sigma'}
  }{\alpha \, {\bar \alpha} \, q_+}.
\end{align}

As the gluon is space-like, $q^2 >0$, the amplitude in \eq{ampl2} has
an imaginary part corresponding to the $q\bar q$ pair in any one of
the fermion bubbles dressing up the gluon line going on mass shell.
However, this imaginary part does not contain any ultraviolet
divergence and hence does not contribute to running coupling
corrections. We will neglect it along with the $5/3$ in the
denominator of \eq{ampl2} and rewrite the amplitude as
\begin{align}\label{ampl3}
  A \, & = \, 4 \, \pi \, i \, \sum\limits_{\lambda = \pm 1} {\bm M}^a
  ({\bm q}, q_+) \cdot {\bm \epsilon}^\lambda \, \frac{\as (q^2)}{q^2}
  \, t^a \, \frac{{\bm \epsilon}^{\lambda *} \cdot {\bm k} \, (1 - 2
    \alpha + \lambda \sigma) \, \delta_{\sigma, \sigma'}}{\alpha \,
    {\bar \alpha} \, q_+}.
\end{align}

Multiplying the amplitude (\ref{ampl3}) by its complex conjugate and
summing over colors, quark helicities $\sigma$, $\sigma'$ and $N_f$
quark flavors (assuming all flavors are massless) yields the amplitude
squared
\begin{align}\label{asq}
  |A|^2 \, = \, (4 \, \pi)^2 \sum\limits_{\lambda, \lambda' = \pm 1} &
  {\bm M}^a ({\bm q}, q_+) \cdot {\bm \epsilon}^\lambda \ {\bm M}^{a
    *} ({\bm q}, q_+) \cdot {\bm \epsilon}^{\lambda' *} \, \left(
    \frac{\as (q^2)}{q^2} \right)^2 \, \frac{N_f}{\alpha^2 \, {\bar
      \alpha}^2 \, q_+^2} \notag \\ & \times \, {\bm
    \epsilon}^{\lambda *} \cdot {\bm k} \ {\bm \epsilon}^{\lambda'}
  \cdot {\bm k} \, [(1 - 2 \alpha)^2 + \lambda \, \lambda' ].
\end{align}

To calculate the contribution to the gluon production cross section of
diagrams with the outgoing gluon line dressed with quark bubbles to
all orders (but excluding the contribution of the bare outgoing gluon
line) we use the identity pictured in \fig{fig:diag_calc}: we flip the sign of the
amplitude squared in \eq{asq} and integrate over $\bm k$ and $\alpha$.
The contribution to the total cross section of the square of the
diagram in \fig{fig:qqbar_ampl} comes with the integration measure
\begin{align}
  \frac{d^2 k_1 \, d k_{1+} \, d^2 k_2 \, d k_{2+}}{2 (2 \pi)^3 \, 2
    (2 \pi)^3} \, = \, \frac{d^2 q \, d^2 k \, d q_{+} \, d \alpha \,
    q_+}{2 (2 \pi)^3 \, 2 (2 \pi)^3}.
\end{align}
Defining the gluon's rapidity $y$ by
\begin{align}\label{gy}
  y \, = \, \frac{1}{2} \, \ln \frac{q_+}{q_-} \, = \, \ln \frac{q_+
    \, \sqrt{2}}{q_T}
\end{align}
with $q_T = |{\bm q}|$ we write the contribution of the diagram on the
left hand side of \fig{fig:diag_calc} to the multiplicity of produced gluons as
\begin{align}\label{NG1}
  \frac{d N^G_\text{bubbles}}{d^2 q \, d y} \, = \, - \int \frac{d^2
    k}{4 (2 \pi)^6} \, \int\limits_0^1 d \alpha \, q_+^2 \, |A|^2.
\end{align}
The integral over $\bm k$ in \eq{NG1} comes with an infrared regulator
$\Lambda_{\text{coll}}$: the domain of integration is restricted to
$k_T > \Lambda_{\text{coll}}$. On the other hand, when arriving at
\eq{ampl1} one effectively integrates over momenta in the (uncut)
quark bubbles up to the ultraviolet (UV) cutoff
$\mu_{\overline{\text{MS}}}$. Therefore, to keep the integration over
$k_T$ in the cut bubble in \fig{fig:qqbar_ampl} in agreement with the
similar integrals in other (uncut) quark bubbles and thus to keep
\fig{fig:diag_calc} valid by restricting all quark bubbles to the
same phase space, we will introduce this UV cutoff and put another
restriction on the range of integration: $k_T <
\mu_{\overline{\text{MS}}}$ . It is important to stress once more that
\eq{NG1} does not contain the contribution of the bare outgoing gluon
line.

Indeed the equality in \fig{fig:diag_calc} is true without any infrared cutoffs on
$k_T$ integration. In other words, the integral over $\bm k$ of the
cut bubble on the right of \fig{fig:diag_calc} is, in general, not restricted to
$k_T > \Lambda_{\text{coll}}$. One can think of the right hand side of
\fig{fig:diag_calc} as consisting of the sum of the $k_T > \Lambda_{\text{coll}}$
and $k_T < \Lambda_{\text{coll}}$ contributions (with an overall minus
sign). In \eq{NG1} we only keep the $k_T > \Lambda_{\text{coll}}$
contribution. Due to \fig{fig:diag_calc} it is equal to the sum of gluon
production and the $q\bar q$ production with $k_T <
\Lambda_{\text{coll}}$. Now the meaning of the multiplicity in
\eq{NG1} becomes clear: it is the net number of produced gluons and
collinear $q\bar q$ pairs with $k_T < \Lambda_{\text{coll}}$. For
simplicity we will refer to the obtained multiplicity distribution as
``gluon'' multiplicity below, but we will always keep in mind that
collinear $q\bar q$ pairs with $k_T < \Lambda_{\text{coll}}$ are
included in it as well.

Replacing $N_f \rightarrow - 6 \pi \beta_2$ in \eq{asq}, plugging the
result into \eq{NG1} we integrate over the angles of $\bm k$ and using
\eq{Ddef} for $q^2$ get
\begin{align}\label{Gmult-1}
  \frac{d N^G_\text{bubbles}}{d^2 q \, d y} \, = \, \frac{6 \, \pi^2
    \, \beta_2}{(2 \, \pi)^4} \ |{\bm M}^a ({\bm q}, q_+)|^2 \,
  \int\limits_0^1 d \alpha \,
  \int\limits_{\Lambda_{\text{coll}}^2}^{\mu_{\overline{\text{MS}}}^2}
  d k_T^2 \, \left( \frac{\as (k_T^2 / \alpha \, {\bar
        \alpha})}{k_T^2} \right)^2 \, \frac{1}{2} \ k_T^2 \, [(1 - 2
  \alpha)^2 + 1].
\end{align}
It is interesting to observe that the argument of the running coupling
in \eq{Gmult-1} is similar to that of the coupling in the
Dokshitzer-Gribov-Lipatov-Altarelli-Parisi (DGLAP) evolution equation
\cite{Dokshitzer:1977sg,Gribov:1972ri,Altarelli:1977zs} as was
calculated in \cite{Dokshitzer:1993pf}.

Using 
\begin{align}\label{ab2}
  \as (Q^2) \, = \, \frac{1}{\beta_2 \, \ln
    \frac{Q^2}{\Lambda_\text{QCD}^2}}
\end{align}
we integrate \eq{Gmult-1} over $k_T^2$. The result of the integration
yields
\begin{align}\label{Gmult1}
  \frac{d N^G_\text{bubbles}}{d^2 q \, d y} \, = \, \frac{3 \,
    \pi^2}{(2 \, \pi)^4} \, |{\bm M}^a ({\bm q}, q_+)|^2 \,
  \int\limits_0^1 d \alpha \, [(1 - 2 \alpha)^2 + 1] \ \left[ \as \!
    \left( \frac{\Lambda_{\text{coll}}^2}{\alpha \, {\bar \alpha}}
    \right) - \as \! \left( \frac{\mu_{\overline{\text{MS}}}^2}{\alpha
        \, {\bar \alpha}} \right) \right].
\end{align}

The part of the $\alpha$-integral in \eq{Gmult1} involving the factor
of $\alpha \, {\bar \alpha}$ in the argument of the coupling constants
contributes only to fixing the constant under the logarithm. In the
spirit of BLM approach \cite{Brodsky:1983gc} we can expand the
coupling constants in \eq{Gmult1} into geometric series, integrate
each (of the first few) terms over $\alpha$, and then resum back the
series. $\alpha \, {\bar \alpha}$ terms in the argument of the
couplings will only contribute constants to the terms in the series:
since we are not keeping track of constant terms we will simply
neglect them. Integrating the prefactor in \eq{Gmult1} over $\alpha$
and neglecting other irrelevant constants in the argument of the
second coupling we write for the contribution to the produced gluon
multiplicity
\begin{align}\label{Gmult2}
  \frac{d N^G_\text{bubbles}}{d^2 q \, d y} \, = \, \frac{1}{(2 \,
    \pi)^2} \, |{\bm M}^a ({\bm q}, q_+)|^2 \ \left[ \as \! \left(
      \Lambda_{\text{coll}}^2 \right) - \am \right].
\end{align}

Finally, adding the contribution of the bare outgoing gluon line
\begin{align}\label{bare}
  \frac{d N^G_\text{bare}}{d^2 q \, d y} \, = \, \frac{1}{2 \, (2 \,
    \pi)^3} \, |{\bm M}^a ({\bm q}, q_+)|^2 \ g^2_\mu \, = \,
  \frac{1}{(2 \, \pi)^2} \, |{\bm M}^a ({\bm q}, q_+)|^2 \ \am
\end{align}
(with $g_\mu$ the bare coupling constant $g$) to the contribution of
\eq{Gmult2} we obtain the following expression for the produced gluon
(and collinear $q\bar q$ pairs) multiplicity with the running coupling
corrections included:
\begin{align}\label{Gmult3}
  \frac{d N^G}{d^2 q \, d y} \, = \, \frac{1}{(2 \, \pi)^2} \, |{\bm
    M}^a ({\bm q}, q_+)|^2 \ \as \! \left( \Lambda_{\text{coll}}^2
  \right).
\end{align}
(Note that \eq{Gmult3} can be obtained directly from \eq{Gmult-1} by
setting the upper limit of the $\bm k$-integration to infinity. As
follows from our derivation above, doing this would modify
\eq{Gmult-1} to include the contribution of the bare gluon line.
Similar observation has been made before for other production
processes in \cite{Beneke:1995pq,Beneke:1998ui}.)

{}From \eq{Gmult3} we can draw the following conclusions:

\begin{itemize}
\item First of all, we have shown that the bare coupling constant
  $\am$ stemming from the vertices of gluon emission by the projectile
  in the amplitude and in the complex conjugate amplitude gets
  renormalized into a physical coupling constant. We have thus
  outlined the procedure of how running coupling corrections should be
  included into the gluon production cross section.
  
\item Unfortunately the scale with which the coupling runs in \eq{Gmult3} is
  some arbitrary infrared cutoff $\Lambda_{\text{coll}}$ which we introduced
  to regulate collinear divergences. This makes the gluon production cross
  section (or, equivalently, gluon multiplicity) either perturbative or
  non-perturbative depending on the choice of the cutoff
  $\Lambda_{\text{coll}}$ corresponding to the resolution scale for gluons and
  $q\bar q$ pairs. Strictly speaking this means that the gluon multiplicity is
  not infrared-safe, in accordance with conventional wisdom.
\end{itemize}

Indeed the dependence of the gluon multiplicity on
$\Lambda_{\text{coll}}$ is already weakened by that fact that it is
absorbed in the argument of a running coupling constant. The
$\Lambda_{\text{coll}}$-dependence could be further reduced by
performing a Sudakov-like resummation \cite{Sudakov:1954sw} of gluon
splittings. If the goal of a calculation is to find the cross section
for production of hadrons, it may be possible to absorb the result of
such resummation into a fragmentation function. However, performing a
Sudakov resummation on top of the calculation of the running coupling
corrections appears to be rather involved and is beyond the scope of
this work.

The resolution scale $\Lambda_{\text{coll}}$ generates an IR cutoff on $k_T$
of the produced quarks in \eq{Gmult-1}. Due to \eq{Ddef} this translates into
a cutoff on gluon virtuality $q^2$, which is equal to the invariant mass of
the $q\bar q$ pair. Neglecting $\alpha \, \bal$ we could say that $q\bar q$
pairs with the invariant mass $q^2 \lesssim \Lambda_{\text{coll}}^2$ are included
into our definition of gluon multiplicity distribution.

However, one should be aware that our calculation is reliable only as long as
$\Lambda_{\text{coll}}^2$ remains in the perturbative domain and that, if one
is forced to take $\Lambda_{\text{coll}}^2$ closer to $\Lambda_{\text{QCD}}^2$
on phenomenological grounds, the only consistent way to take into account the
non-perturbative contributions -- and the nontrivial true dependence on
$\Lambda_{\text{coll}}^2$ in the non-perturbative region -- is to introduce a
factorization procedure that subsumes these contributions in a fragmentation
function whose $\Lambda_{\text{coll}}^2$ dependence can only be inferred with
non-perturbative means or by comparison with experiment.  This introduces a
factorization scale $\mu^2$ between perturbative contributions above it and
non-perturbative ones below.  The former are reliably reproduced by our
calculation if we substitute the perturbative factorization scale $\mu^2$ in
place of $\Lambda_{\text{coll}}^2$ in the above.\footnote{In such a setting,
  one has to keep in mind that the factorization scale, while infrared from
  the perspective of the perturbative contribution, is in the UV from the
  perspective of the modes entering the operator defining the fragmentation
  functions \cite{Berger:1996vy,Berger:1995fm} and may be thought of as its
  renormalization scale $\mu$, which, in general, is different from full UV
  renormalization scale $\mu_{\overline{\text{MS}}}$ for the complete graph.
  Fragmentation functions depend both on the factorization scale $\mu$ and on
  the collinear IR cutoff $\Lambda_{\text{coll}}$. The perturbative part of
  the cross section depends on the renormalization scale
  $\mu_{\overline{\text{MS}}}$ and on the factorization scale $\mu$. In the
  full hadron production cross section the $\mu$-dependence cancels so that it
  only depends on $\mu_{\overline{\text{MS}}}$ and $\Lambda_{\text{coll}}$.}
For heavy quark--anti-quark bound state production the factorization scale is
chosen to be equal to the lowest possible value of the invariant mass of the
$q\bar q$ pair \cite{Braaten:1993rw}, which is not too different from the
cutoff on the invariant mass obtained above. Still, the introduction of
fragmentation functions is necessary for consistency even in such a case
where the non-perturbative contributions become (almost) trivial.

In practical phenomenological calculations of hadron production one
often chooses the factorization scale $\mu \sim p_T$ with $p_T$ the
transverse momentum of the produced hadron. In this case, \eq{Gmult3}
should be convoluted with the fragmentation functions and
$\Lambda_\text{coll}$ in it should be replaced by the factorization
scale $\mu$.  If one puts $\mu \sim p_T$ one would then get $\as
(p_T^2)$ in \eq{Gmult3}. While in an all-order calculation the hadron
production cross section should not depend on the choice of the
factorization scale $\mu$, in practical applications putting $\mu \sim
p_T$ in our result would lead to some additional $p_T$-dependence.



\section{Collinear Singularities and the Energy Density}
\label{energy}

Our goal now is to verify that while gluon (and quark) multiplicity is not
infrared-finite, the energy density of the medium produced in the collision is
infrared-finite. The energy density $\epsilon$ is given by \eq{e}. With the
gluon multiplicity being infinite, there are at least two ways the net energy
density could stay finite:
\begin{itemize}
\item[(i)] it may be that the sum of gluon and quark multiplicities
  (net parton multiplicity) is finite, or
\item[(ii)] it may be that the divergence in the sum of gluon and
  quark multiplicities vanishes after the integration over the
  transverse momentum in \eq{e}.
\end{itemize}

To calculate the energy density using \eq{e} we need the expressions
for gluon and quark multiplicities. For the gluon multiplicity we have
\eq{Gmult3}. To calculate the contribution of the diagram in \fig{fig:qqbar_ampl}
to the quark production cross section, we simply have to take the
expression in \eq{Gmult-1}, flip its sign back (to undo the sign flip
shown on the right hand side of \fig{fig:diag_calc}) and instead of integrating
over $\bm k$ and $\alpha$ we would integrate over momenta of the
anti-quark keeping the transverse momentum and rapidity of the quark
fixed. We get
\begin{align}\label{qmult1}
  \frac{d N^q}{d^2 p \, d y} \, = \, - \frac{1}{(2 \, \pi)^4} \int d^2
  k_1 \, d^2 k_2 \, \frac{d k_{1+} \, d k_{2+}}{q_+^2} \, \delta({\bm
    p} - {\bm k}_1) \, \delta \! \left( y - \ln \frac{k_{1+} \,
      \sqrt{2}}{k_{1T}} \right) \, |{\bm M}^a ({\bm q}, q_+)|^2 \, f
  ({\bm k}; \alpha, {\bar \alpha})
\end{align}
were we have defined
\begin{align}\label{f}
  f ({\bm k}; \alpha, {\bar \alpha}) \, = \, 6 \, \pi \, \beta_2 \,
  \left( \frac{\as (k_T^2 / \alpha \, {\bar \alpha})}{k_T^2} \right)^2
  \, \frac{1}{2} \ k_T^2 \, [(\alpha - {\bar \alpha})^2 + 1].
\end{align}
The produced quark's rapidity is given by $y = \ln (p_+ \, \sqrt{2} /
p_T)$, same as \eq{gy} for the gluon rapidity. Note that $f ({\bm k};
\alpha, {\bar \alpha})$ is symmetric under the interchange $\alpha
\leftrightarrow \bal$: $f ({\bm k}; \alpha, {\bar \alpha}) = f ({\bm
  k}; \bal, \alpha)$.

Strictly speaking to calculate quark production we would have to
replace $- 6 \, \pi \, \beta_2$ in \eq{f} (and, consequently, in
\eq{qmult1}) back with $N_f$. However we will leave $\beta_2$ intact
in \eq{f}. By doing so we will indeed include the cut gluon bubble
into \eq{qmult1} making the expression in \eq{qmult1} include both
quark multiplicity and the multiplicity of gluons coming from a cut
gluon bubble correction. Indeed if we want to see cancellation or
disappearance of collinear singularities, we need to sum both
(anti-)quark and gluon contributions. Hence, in our notation, the
"quark multiplicity" would include both quark and gluon multiplicity
coming from the cut bubbles.

Eliminating the delta-functions in \eq{qmult1} we write
\begin{align}\label{qmult2}
  \frac{d N^q}{d^2 p \, d y} \, = \, - \frac{1}{(2 \, \pi)^4} \int d^2
  k_2 \, \int\limits_0^1 \, d \alpha \, \bigg| {\bm M}^a \left({\bm p}
    + {\bm k}_2, \frac{p_T \, e^y}{\alpha \, \sqrt{2}} \right)
  \bigg|^2 \, f ({\bar \alpha} \, {\bm p} - \alpha \, {\bm k}_2;
  \alpha, {\bar \alpha})
\end{align}
were, in accordance with \eq{alpha}, we have used
\begin{align}
\alpha \, = \, \frac{p_+}{p_+ + k_{2+}}.
\end{align}

The anti-quark multiplicity is obtained from the multiplicity of the
quarks in (\ref{qmult2}) by replacing $\alpha \leftrightarrow {\bar
  \alpha}$ and ${\bm k}_2 \rightarrow {\bm k}_1$. We get
\begin{align}\label{aqmult}
  \frac{d N^{\bar q}}{d^2 p \, d y} \, = \, - \frac{1}{(2 \, \pi)^4}
  \int d^2 k_1 \, \int\limits_0^1 \, d \alpha \, \bigg| {\bm M}^a
  \left({\bm p} + {\bm k}_1, \frac{p_T \, e^y}{\bal \, \sqrt{2}}
  \right) \bigg|^2 \, f ({\alpha} \, {\bm p} - \bal \, {\bm k}_1;
  \alpha, {\bar \alpha}).
\end{align}

With the help of Eqs. (\ref{Gmult-1}) and (\ref{f}) we rewrite
\eq{Gmult3} as
\begin{align}\label{Gmult}
  \frac{d N^{G}}{d^2 p \, d y} \, = \, \frac{1}{(2 \, \pi)^4} \int d^2
  k \, \int\limits_0^1 \, d \alpha \, \bigg| {\bm M}^a \left({\bm p},
    \frac{p_T \, e^y}{\sqrt{2}} \right) \bigg|^2 \, f ({\bm k};
  \alpha, {\bar \alpha})
\end{align}
where now the integral over $k_T$ is not bounded by the cutoff
$\mu_{\overline {\text{MS}}}$ from above. As was discussed in the
previous section this insures that the contribution of the bare
outgoing gluon line is included in \eq{Gmult} as well.

Defining the net parton multiplicity as the sum of quark and gluon
contributions
\begin{align}\label{Pmult_def}
  \frac{d N^{\text{partons}}}{d^2 p \, d y} \, := \, \frac{d N^{G}}{d^2
    p \, d y} + \frac{d N^q}{d^2 p \, d y} + \frac{d N^{\bar q}}{d^2 p
    \, d y}
\end{align}
we write, using Eqs. (\ref{qmult2}), (\ref{aqmult}) and (\ref{Gmult})
\begin{align}\label{Pmult}
  & \frac{d N^{\text{partons}}}{d^2 p \, d y} \, = \, \frac{1}{(2 \,
    \pi)^4} \int d^2 k \, \int\limits_0^1 \, d \alpha \, \Bigg\{
  \bigg| {\bm M}^a \left({\bm p}, \frac{p_T \, e^y}{\sqrt{2}} \right)
  \bigg|^2 \, f ({\bm k}; \alpha,  {\bar \alpha}) \notag \\
  & - \bigg| {\bm M}^a \left({\bm p} + {\bm k}, \frac{p_T \,
      e^y}{\alpha \, \sqrt{2}} \right) \bigg|^2 \, f ({\bar \alpha} \,
  {\bm p} - \alpha \, {\bm k}; \alpha, {\bar \alpha}) - \bigg| {\bm
    M}^a \left({\bm p} + {\bm k}, \frac{p_T \, e^y}{\bal \, \sqrt{2}}
  \right) \bigg|^2 \, f ({\alpha} \, {\bm p} - \bal \, {\bm k};
  \alpha, {\bar \alpha}) \Bigg\}
\ ,
\end{align}
where we have relabeled the integration variables ${\bm k}_1$ and
${\bm k}_2$ in Eqs. (\ref{aqmult}) and (\ref{qmult2}) as $\bm k$.

To analyze collinear divergences in \eq{Pmult} it is convenient to
redefine transverse momenta in the arguments of the functions $f$ in
the second and the third term in the curly brackets in \eq{Pmult}. We
replace $\alpha \, {\bm k} - {\bar \alpha} \, {\bm p} \rightarrow {\bm
  k}$ in the second term and $\bal \, {\bm k} - {\alpha} \, {\bm p}
\rightarrow {\bm k}$ in the third term to obtain
\begin{align}\label{Pmult1}
  \frac{d N^{\text{partons}}}{d^2 p \, d y} \, = \, & \frac{1}{(2 \,
    \pi)^4} \int d^2 k \, \int\limits_0^1 \, d \alpha \ f ({\bm k};
  \alpha, {\bar \alpha}) \ \Bigg\{ \bigg| {\bm M}^a \left({\bm p},
    \frac{p_T \, e^y}{\sqrt{2}} \right)
  \bigg|^2 \,  \notag \\
  & - \frac{1}{\alpha^2} \, \bigg| {\bm M}^a \left(\frac{{\bm p} +
      {\bm k}}{\alpha}, \frac{p_T \, e^y}{\alpha \, \sqrt{2}} \right)
  \bigg|^2 - \frac{1}{\bal^2} \, \bigg| {\bm M}^a \left(\frac{{\bm p}
      + {\bm k}}{\bal}, \frac{p_T \, e^y}{\bal \, \sqrt{2}} \right)
  \bigg|^2 \Bigg\}.
\end{align}
With the help of \eq{Pmult1} we clearly see that the collinear
divergence at ${\bm k} = 0$ does not get canceled between the terms in
the curly brackets. 

One might imagine some special ``fine tuning'' choices of amplitudes
$\bm M$ (e.g. 
\begin{align}
\left|{\bm M} \left({\bm p}, \frac{p_T \, e^y}{\sqrt{2}}
\right)\right|^2 \, \sim \, 1/p_T^2 \notag
\end{align}
without any rapidity dependence) for
which such cancellation would be possible: however, even at the lowest
order in the coupling one has \cite{Kovner:1995ja,Kovchegov:1997ke}
$\left|{\bm M} \left({\bm p}, \frac{p_T \, e^y}{\sqrt{2}} \right)\right|^2 \,
\sim \, 1/p_T^4$ and cancellation does not happen in \eq{Pmult1}.
Higher order rescatterings or other corrections are not likely to
modify this scaling of the amplitude squared for sufficiently high
$p_T$. Hence the special choice of $\left|{\bm M} \left({\bm p}, \frac{p_T
    \, e^y}{\sqrt{2}} \right)\right|^2 \, \sim \, 1/p_T^2$ for all $p_T$
appears to be not achievable in QCD.

Therefore, the possibility (i) outlined above is ruled out: the net
parton multiplicity still contains collinear divergences. 

Before we proceed it is important to point out that, while similar to
\eq{Gmult-1} the $\bm k$-integrals in the second and third (quark and
anti-quark) terms in the curly brackets in \eq{Pmult1} are cut off by
$\mu_{\overline{\text{MS}}}$ in the ultraviolet, the $\bm k$-integral
in the first (gluon) term in the curly brackets is not bounded in the
UV. This is done to include the contribution of the bare outgoing
gluon line, as was discussed in the previous Section and noted again
in this Section after \eq{Gmult}. Indeed the UV behavior of the
integrals is not essential for our conclusion of non-cancellation of
collinear divergences in \eq{Pmult1}, but it will help us understand
the physical picture later on.

To verify the cancellation of collinear divergences in the energy
density, we first combine \eq{e} and \eq{Pmult_def} to write
\begin{align}\label{e_part}
  \epsilon (\tau, \eta, {\bm b}) \, = \, \frac{1}{\tau \, S_\perp} \,
  \int d^2 p \, p_T \ \frac{d N^\text{partons}}{d^2 p \, d \eta}.
\end{align}
In arriving at \eq{e_part} we have assumed the target (nucleus) to be
cylindrical oriented along the collision axis. This allowed us to
simplify the calculation by replacing $d^2 b$ in the denominator by
the transverse cross-sectional area of the nucleus $S_\perp$. This
simplification does not affect collinear divergences. 

Substituting \eq{Pmult1} into \eq{e_part} we write
\begin{align}\label{e1}
  \epsilon (\tau, \eta, {\bm b}) \, = \, \frac{1}{\tau \, S_\perp \,
    (2 \, \pi)^4} \, \int d^2 p \, p_T \ \int d^2 k \, \int\limits_0^1
  \, d \alpha \ f ({\bm k}; \alpha, {\bar \alpha}) \ \Bigg\{ \bigg|
  {\bm M}^a \left({\bm p}, \frac{p_T \, e^\eta}{\sqrt{2}} \right)
  \bigg|^2 \,  \notag \\
  - \frac{1}{\alpha^2} \, \bigg| {\bm M}^a \left(\frac{{\bm p} + {\bm
        k}}{\alpha}, \frac{p_T \, e^\eta}{\alpha \, \sqrt{2}} \right)
  \bigg|^2 - \frac{1}{\bal^2} \, \bigg| {\bm M}^a \left(\frac{{\bm p}
      + {\bm k}}{\bal}, \frac{p_T \, e^\eta}{\bal \, \sqrt{2}} \right)
  \bigg|^2 \Bigg\}.
\end{align}
Again, the collinear singularity is at ${\bm k} =0$. Near the
singularity we can put ${\bm k} =0$ in the arguments of $|{\bm M}|^2$
in the second and the third terms in the curly brackets of \eq{e1}.
The (potentially) singular part of the energy density is
\begin{align}\label{e_sing}
  \epsilon_\text{sing} \, = \, \frac{1}{\tau \, S_\perp \, (2 \,
    \pi)^4} \, \int d^2 p \ p_T \ \int d^2 k \, \int\limits_0^1 \, d
  \alpha \ f ({\bm k}; \alpha, {\bar \alpha}) \ \Bigg\{ \bigg| {\bm
    M}^a \left({\bm p}, \frac{p_T \, e^\eta}{\sqrt{2}} \right)
  \bigg|^2 \,  \notag \\
  - \frac{1}{\alpha^2} \, \bigg| {\bm M}^a \left(\frac{{\bm
        p}}{\alpha}, \frac{p_T \, e^\eta}{\alpha \, \sqrt{2}} \right)
  \bigg|^2 - \frac{1}{\bal^2} \, \bigg| {\bm M}^a \left(\frac{{\bm
        p}}{\bal}, \frac{p_T \, e^\eta}{\bal \, \sqrt{2}} \right)
  \bigg|^2 \Bigg\}.
\end{align}
Rescaling $p_T$ by $\alpha$ in the second term in the curly brackets
and by $\bal$ in the third term in the curly brackets in \eq{e_sing}
we see that the collinear singularities cancel in the energy density!

Therefore, the option (ii) outlined above is correct! Energy density is
independent of collinear splittings, while parton multiplicity is not.

Let us point out that, as was observed before, the $k_T$-integrals in
the second and third terms in \eq{e_sing} have a UV cutoff
$\mu_{\overline{\text{MS}}}$, while the first term is not limited in
the ultraviolet. Therefore, strictly speaking \eq{e_sing} gives
\begin{align}\label{e_bare1}
  \epsilon_\text{sing} \, = \, \frac{\pi}{\tau \, S_\perp \, (2 \,
    \pi)^4} \, \int d^2 p \ p_T \ 
  \int\limits_{\mu_{\overline{\text{MS}}}^2}^\infty d k_T^2 \,
  \int\limits_0^1 \, d \alpha \ f ({\bm k}; \alpha, {\bar \alpha}) \ 
  \bigg| {\bm M}^a \left({\bm p}, \frac{p_T \, e^\eta}{\sqrt{2}}
  \right) \bigg|^2.
\end{align}
(For simplicity again we do not distinguish between
$\mu_{\overline{\text{MS}}}$ and, say,
$\mu_{\overline{\text{MS}}}/\alpha$.)  While the term in \eq{e_bare1}
has no collinear divergences with $k_T$ integral running over the UV
modes, it is nevertheless non-zero. Integrating over $k_T$ and
$\alpha$ we obtain
\begin{align}\label{e_bare}
  \epsilon_\text{sing} \, = \, \frac{1}{\tau \, S_\perp \, (2 \,
    \pi)^2} \, \int d^2 p \ p_T \ \am \, \bigg| {\bm M}^a \left({\bm
      p}, \frac{p_T \, e^\eta}{\sqrt{2}} \right) \bigg|^2.
\end{align}
This is indeed the contribution of the bare outgoing gluon line to the
energy density. 

The physical picture behind the transition from \eq{e_sing} to
\eq{e_bare} is clear. When a gluon splits into a collinear $q\bar q$
pair (or into a pair of gluons) the energy density deposited into a
region of space would not change (in the strictly collinear limit).
Therefore, energy density is not affected by collinear
divergences/splittings. What remains of the energy density
(\ref{e_sing}), which included all real and virtual splitting, is the
energy of the original gluon in \eq{e_bare}, as it should be: none of
the collinear splittings would change the energy density.

\begin{figure}[htb]
  \centering
\parbox{7cm}{\includegraphics[width=7cm]{taggedglue}}
\ \ +\ \ \parbox{7cm}{\includegraphics[width=7cm]{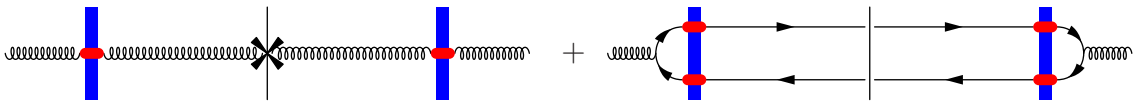}}
  \caption{Addition of the diagram with the quark bubble (on the right) 
    needed to renormalize the energy density $\epsilon$ discussed in
    the text. The diagram on the right was also needed in
    \protect\cite{Balitsky:2006wa,Kovchegov:2006vj} to take into
    account the running coupling corrections to the total cross
    section, as calculated using small-$x$ JIMWLK and BK evolution
    equations. }
  \label{fig:erenorm}
\end{figure}

Of course one may be worried that the energy density in \eq{e_bare}
contains a bare coupling constant $\am$ instead of a physical coupling
$\as$. This bare coupling gets renormalized in a manner similar to
what was done in~\cite{Balitsky:2006wa,Kovchegov:2006vj} for the total
scattering cross section: one has to add to the bare gluon term a
diagram with a quark bubble (to be completed to the full beta-function
by a similar gluon bubble along with other gluonic corrections), where
the quark bubble both begins and ends in the initial state. This is
shown in \fig{fig:erenorm}.
As was demonstrated in \cite{Balitsky:2006wa,Kovchegov:2006vj} the
quark bubble diagram contains a UV divergence coming from the region
of the transverse coordinate integral where the quark and the
anti-quark are on top of each other. This divergence was instrumental
in renormalizing the total cross section in
\cite{Balitsky:2006wa,Kovchegov:2006vj}, and is likely to renormalize
the energy density in \eq{e_bare}.


\section{Toy Model}
\label{toy_sect}

In this Section we will use a toy model for the amplitude squared to
illustrate how the collinearly divergent expression for the net parton
multiplicity (\ref{Pmult1}) leads to a finite energy density after the
integration in \eq{e_part}. Using the same approximations and variable
redefinitions that led to \eq{e_sing} we write the singular parts of
the gluon and quark spectra as
\begin{align}\label{Gtoy}
  \frac{d N^{G}_\text{sing}}{d^2 p \, d y} \, = \, \frac{1}{(2 \,
    \pi)^4} \int d^2 k \, \int\limits_0^1 \, d \alpha \, \bigg| {\bm
    M}^a \left({\bm p}, \frac{p_T \, e^y}{\sqrt{2}} \right) \bigg|^2
  \, f ({\bm k}; \alpha, {\bar \alpha}) \, = \, \frac{1}{(2 \, \pi)^2}
  \, |{\bm M}^a ({\bm p}, p_+)|^2 \ \as \! \left(
    \Lambda_{\text{coll}}^2 \right),
\end{align}
\begin{align}\label{qtoy}
  \frac{d N^{q}_\text{sing}}{d^2 p \, d y} \, = \, - \frac{1}{(2 \,
    \pi)^4} \int d^2 k \, \int\limits_0^1 \, d \alpha \,
  \frac{1}{\alpha^2} \, \bigg| {\bm M}^a \left(\frac{{\bm p}}{\alpha},
    \frac{p_T \, e^y}{\alpha \, \sqrt{2}} \right) \bigg|^2 \, f ({\bm
    k}; \alpha, {\bar \alpha}),
\end{align}
\begin{align}\label{aqtoy}
  \frac{d N^{\bar q}_\text{sing}}{d^2 p \, d y} \, = \, - \frac{1}{(2
    \, \pi)^4} \int d^2 k \, \int\limits_0^1 \, d \alpha \,
  \frac{1}{\bal^2} \, \bigg| {\bm M}^a \left(\frac{{\bm p}}{\bal},
    \frac{p_T \, e^y}{\bal \, \sqrt{2}} \right) \bigg|^2 \, f ({\bm
    k}; \alpha, {\bar \alpha}).
\end{align}

For simplicity let us write 
\begin{align}\label{ff}
  f ({\bm k}; \alpha, {\bar \alpha}) \, = \, 6 \, \pi \, \beta_2 \,
  \left( \frac{\as (k_T^2)}{k_T^2} \right)^2 \, \frac{1}{2} \ k_T^2 \,
  [(\alpha - {\bar \alpha})^2 + 1],
\end{align}
which is obtained from \eq{f} by dropping $\alpha \, \bal$ in the
argument of the coupling constant. Indeed, in all of the above
calculations that factor was neglected in the end: we might as well
neglect it right away here.

Using \eq{ff} in Eqs. (\ref{qtoy}) and (\ref{aqtoy}) we perform $k_T$
integrals (remembering that $\Lambda_\text{coll} < k_T <
\mu_{\overline{\text{MS}}}$) obtaining
\begin{align}\label{qtoy1}
  \frac{d N^{q}_\text{sing}}{d^2 p \, d y} \, = \, \frac{d N^{\bar
      q}_\text{sing}}{d^2 p \, d y} \, = \, - \frac{3 \, \pi^2}{(2 \,
    \pi)^4} \, \int\limits_0^1 \, d \alpha \, \frac{1}{\alpha^2} \,
  [(\alpha - {\bar \alpha})^2 + 1] \, \bigg| {\bm M}^a
  \left(\frac{{\bm p}}{\alpha}, \frac{p_T \, e^y}{\alpha \, \sqrt{2}}
  \right) \bigg|^2 \, \left[ \as \! \left( \Lambda_{\text{coll}}^2
    \right) - \am \right].
\end{align}

Now, as a toy model, and in the spirit of saturation approach let us
take
\begin{align}\label{Mtoy}
  \frac{1}{(2 \, \pi)^2} \, |{\bm M}^a ({\bm p}, p_+)|^2 = \left\{
    \begin{matrix}
      \frac{1}{p_T^4} \ , p_T > Q_s \cr \cr \frac{1}{Q_s^4} \ , p_T <
      Q_s \cr
\end{matrix}
\right.
\end{align}
with some proportionality coefficient, which we put to one for
simplicity.

With the amplitude squared given by \eq{Mtoy} substituted into
\eq{Gtoy} we get
\begin{align}\label{Gtoy2}
  \frac{d N^{G}_\text{sing}}{d^2 p \, d y} \, = \, \as \! \left(
    \Lambda_{\text{coll}}^2 \right) \, \left\{
    \begin{matrix}
      \frac{1}{p_T^4} \ , p_T > Q_s \cr \cr \frac{1}{Q_s^4} \ , p_T <
      Q_s \cr
\end{matrix}
\right.
\end{align}
for the number of produced gluons.

Substituting \eq{Mtoy} into \eq{qtoy1} and integrating yields
\begin{align}\label{qtoy2}
  \frac{d N^{q}_\text{sing}}{d^2 p \, d y} \, = \, \frac{d N^{\bar
      q}_\text{sing}}{d^2 p \, d y} \, = \, -  \left[
    \as \! \left( \Lambda_{\text{coll}}^2 \right) - \am \right] \,
  \left\{
    \begin{matrix}
      \frac{7}{20 \, p_T^4} \ \ , p_T > Q_s \cr \cr \frac{2}{p_T \,
        Q_s^3} + \frac{3}{4 \, Q_s^4} - \frac{3}{Q_s^4} \, \ln
      \frac{Q_s}{p_T} - \frac{12 \, p_T}{5 \, Q_s^5} \ \ , p_T < Q_s .
      \cr
\end{matrix}
\right. 
\end{align}
The term containing $\am$ in \eq{qtoy2} is the contribution of the
bare gluon, which is independent of the collinear cutoff
$\Lambda_{\text{coll}}$. Therefore, since we are interested only in
the cancellation mechanism for collinear divergences, we will drop
this term in the further calculations.

Using Eqs. (\ref{Gtoy2}) and (\ref{qtoy2}) in \eq{Pmult_def} we obtain
the $\Lambda_{\text{coll}}$-dependent part of the net parton
multiplicity
\begin{align}\label{ptoy}
  \frac{d N^{\text{partons}}_\text{sing}}{d^2 p \, d y} \, = \, \as \!
  \left( \Lambda_{\text{coll}}^2 \right) \, \left\{
    \begin{matrix}
      \frac{3}{10 \, p_T^4} \ \ , p_T > Q_s \cr \cr - \frac{4}{p_T \,
        Q_s^3} - \frac{1}{2 \, Q_s^4} + \frac{6}{Q_s^4} \, \ln
      \frac{Q_s}{p_T} + \frac{24 \, p_T}{5 \, Q_s^5} \ \ , p_T < Q_s .
      \cr
\end{matrix}
\right.
\end{align}
Now one can easily see that plugging \eq{ptoy} in \eq{e_part} and
integrating over $p_T$ would give zero contribution to the energy
density: 
\begin{align}
  \int d^2 p \, p_T \, \frac{d N^{\text{partons}}_\text{sing}}{d^2 p
    \, d y} \, = \, 0.
\end{align}
The collinearly divergent piece given in \eq{ptoy} integrates out to
zero! This is indeed due to the fact that $d
N^{\text{partons}}_\text{sing}/d^2 p \, d y$ is {\sl negative} for
some values of $p_T$ in the range of $p_T < Q_s$.

What we have learned from the above toy model is that the collinearly
divergent part of the net parton multiplicity $d
N^{\text{partons}}/d^2 p \, d y$, while non-zero, comes with a
$p_T$-dependent coefficient, which integrates out to zero when weighed
with $p_T$ and integrated over $p_T^2$ to obtain the energy density
using \eq{e_part}. This clarifies the question of how exactly the
energy density remains finite while $d N^{\text{partons}}/d^2 p \, d
y$ is not.


\section{Outlook}
\label{out}

In this paper we have outlined the inclusion of running coupling
corrections into the multiplicity distribution of the produced gluons
$d N^G/d^2 p \, d y$ for the scattering of a small projectile on a
larger target. We found that both the running coupling corrections and
the collinear singularities enter at the same order in the coupling.
Therefore, inevitably the gluon multiplicity distribution turns out to
be dependent on the infrared cutoff $\Lambda_\text{coll}$ which we
inserted to regulate the collinear divergences. This result is indeed
in agreement with the conventional understanding of gluon (or any
other massless parton) production. However, when we resumed running
coupling corrections (defined as powers of $\am \, N_f$ with $N_f$
later completed to the full beta-function) to all orders, it turned
out that the IR cutoff $\Lambda_\text{coll}$ sets the scale of one of
the factors of the running coupling constant, as shown in \eq{Gmult3}.
This would indeed make gluon multiplicity distribution not infrared
safe. For hadron production calculations one would expect that the
contribution of the collinear singularities could be factored out into
the fragmentation function.

We have then proceeded to analyze the effect of collinear
singularities on the energy density of the medium produced in the
collision as defined by \eq{e} above (see \cite{Kovchegov:2005ss}). In
Sect. \ref{energy} we show that the energy density $\epsilon$ is
independent of $\Lambda_\text{coll}$. Therefore the energy density is
infrared safe.  This result is easy to interpret physically, as
collinear splittings, where a gluon splits into two gluons (or into a
$q\bar q$ pair) both of which are flying in the same direction, are
not going to change the amount of energy density deposited into a
given region of space.  We have also clarified how the parton
multiplicity in \eq{e} could be divergent due to collinear
singularities but give a finite energy density after $p_T$ weighing
and integration in \eq{e}.  It turns out, as was shown in Sect.
\ref{toy_sect}, that the collinearly divergent part of the parton
multiplicity comes in multiplied by a function of $p_T$ which, when
weighed with $p_T$ and integrated over all $p_T^2$, integrates out to
zero. It is interesting to note that the energy density in \eq{e} is
proportional to another infrared-safe quantity called the energy flow,
which was defined in \cite{Kwiecinski:1994zs} as a potential signal
for the experimental detection of the BFKL evolution in deep inelastic
scattering (DIS).

Collinear singularities modify the proof presented in \cite{Kovchegov:2005ss}
of the impossibility of perturbative isotropization (and thermalization) in
heavy ion collisions which led to \eq{e}. To account for the collinear
singularities one has to separate each function $f_i (p^2, p \cdot p', p'^2,
p_T)$ employed in \cite{Kovchegov:2005ss} (with $k$ used in place of $p$ in
\cite{Kovchegov:2005ss}) into collinearly divergent and finite at $p^2 = 0$
(or $p'^2 =0$) parts.  For the singular part one has to introduce an IR
regulator $\Lambda_\text{coll}$ as a mass in the denominator of the propagator
of the outgoing gluon (or quark for the part of the argument made for the
quarks). As was shown above in Sections \ref{energy} and \ref{toy_sect}, the
terms in $\epsilon$ which are singular in the $\Lambda_\text{coll} \rightarrow
0$ limit will vanish after $p_T$-integration. For the remaining
collinearly-finite part of the function $f_i (p^2, p \cdot p', p'^2, p_T)$ the
proof continues as was outlined in \cite{Kovchegov:2005ss}.  The conclusion of
no perturbative isotropization is therefore not affected by collinear
singularities.

While in this paper we have outlined the inclusion of running coupling
corrections into the gluon production cross section, the full
calculation still remains to be performed. It would indeed be
important for improving our understanding of which part of the
contributions can be expected to be made infrared safe by the presence
of the saturation scale. As long as the other contributions can be
treated with factorization methods, albeit at the price of having to
introduce non-perturbative quantities like fragmentation functions, one
arrives at a perturbatively consistent calculational framework.  In
such a framework, knowledge of the scales of the running couplings in
the unintegrated gluon distribution functions which enter the
$k_T$-factorization formula for gluon production in \eq{ktfact}
\cite{Gribov:1981ac,Gribov:1984tu}, which was shown to be valid for
gluon production in $pA$ collisions and in DIS
\cite{Kovchegov:2001sc,Kovchegov:1998bi,Kovchegov:2005ur}, is of
paramount importance.  It would allow one to make predictions for
hadronic spectra and multiplicities in nuclear and hadronic collisions
based on CGC physics with a much higher precision than was ever
possible before. Such results would be extremely valuable for the
upcoming LHC and EIC/eRHIC experimental programs.



\section*{Acknowledgments} 

Yu.K. would like to thank Francois Gelis, discussions with whom during
the University of Washington's Institute for Nuclear Theory program
``From RHIC to LHC: Achievements and Opportunities'' got Yu.K.
interested in the subject discussed in this paper. The authors would
like to thank Ian Balitsky, Eric Braaten, Al Mueller and Jianwei Qiu
for interesting and helpful comments on the problem at hand.

The research of Yu.K. is sponsored in part by the U.S. Department of
Energy under Grant No. DE-FG02-05ER41377. H.W. is supported by the
University of Oulu.


\appendix

\renewcommand{\theequation}{A\arabic{equation}}
  \setcounter{equation}{0}
\section{Running Coupling Corrections for Mueller-Glauber Multiple Rescatterings}
\label{p-np}

\begin{figure}[htb]
  \centering
    \includegraphics[width=7cm]{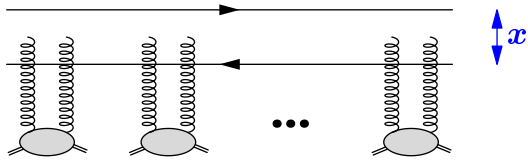}
  \caption{Glauber-Mueller multiple rescatterings of a quark-antiquark 
    dipole on a nuclear target.}
  \label{fig:mugla}
\end{figure}

Here we will discuss running coupling corrections to multiple
exchanges of $t$-channel gluons. We will consider scattering of a
$q\bar q$ dipole on the nucleus in the quasi-classical approximation
with two-gluon exchange interactions between the dipole and each
nucleon, as was first calculated in \cite{Mueller:1989st}. This
approximation is equivalent to the classical McLerran-Venugopalan
model
\cite{McLerran:1993ni,McLerran:1993ka,McLerran:1994vd,Kovchegov:1996ty,Kovchegov:1997pc,Jalilian-Marian:1997xn}.
To concentrate on the running coupling corrections to the $t$-channel
gluons let us take a quark-antiquark dipole as a projectile. The
forward amplitude for multiple rescatterings resummed in
\cite{Mueller:1989st} is shown in \fig{fig:mugla}.
As was demonstrated in \cite{Mueller:1989st} a sum of such multiple
rescattering diagrams gives the forward scattering amplitude
\begin{align}\label{N0}
  N ({\bm x}, {\bm b}) \, = \, 1 - \exp \left( - \pi \, \am^2 \,
    \frac{C_F}{N_c} \, \rho \, T({\bm b}) \, {\bm x}^2 \, \ln
    \frac{1}{|\bm x| \, \Lambda} \right).
\end{align}
The above amplitude is for a dipole of transverse size ${\bm x}$
scattering on the target nucleus at the impact parameter $\bm b$.
$\rho$ is the density of nucleons in the nucleus, $T({\bm b})$ is the
nuclear profile function equal to the length of the nuclear medium at
the impact parameter $\bm b$, such that $T({\bm b}) = 2 \, \sqrt{R^2 -
  b^2}$ for a spherical nucleus of radius $R$. Also, $x_\perp = |{\bm
  x}|$, $C_F = \frac{N_c^2 - 1}{2 \, N_c}$ and $\Lambda$ is the
(non-perturbative) momentum scale characterizing each nucleon. We have
put $\am$ as the bare coupling constant to underline that the
calculation of \cite{Mueller:1989st} leading to \eq{N0} was done for
fixed coupling. Until the running coupling corrections are included
the coupling in \eq{N0} is the bare coupling.

One can show (see \cite{Jalilian-Marian:2005jf} for a pedagogical
derivation) that the expression in the exponent of \eq{N0} is equal to
\begin{align}
  - \frac{\rho \, T({\bm b}) \, \sigma^{q{\bar q} N}}{2}
\end{align}
with $\sigma^{q{\bar q} N}$ the scattering cross section of the dipole
on a single nucleon. The dipole-nucleon cross section due to a
two-gluon exchange is given by
\begin{align}\label{sigN}
  \sigma^{q{\bar q} N} \, = \, 2 \, \am^2 \, \frac{C_F}{N_c} \, \int
  \frac{d^2 l}{[{\bm l}^2]^2} \, \left( 2 - e^{i {\bm l} \cdot {\bm
        x}} - e^{- i {\bm l} \cdot {\bm x}} \right) \, = \, 2 \, \am^2
  \, \frac{C_F}{N_c} \, \pi \, {\bm x}^2 \, \ln \frac{1}{|\bm x| \,
    \Lambda},
\end{align}
where the two factors of ${\bm l}^2$ in the denominator come from the
propagators of the two gluons. $\Lambda$ is used as the infrared
cutoff of the $l$-integral in \eq{sigN}.

\begin{figure}[htb]
  \centering
  \includegraphics[width=4cm]{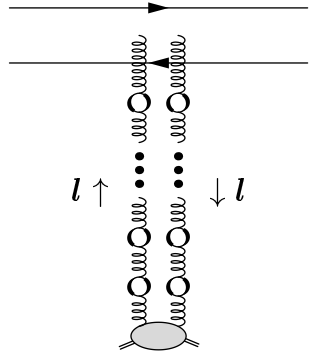}
  \caption{The interaction of a $q\bar q$ dipole with a single nucleon 
    in the nucleus in the Glauber-Mueller approximation ``dressed''
    with quark bubble corrections.}
  \label{fig:dress}
\end{figure}

To include running coupling corrections into \eq{N0}, it is sufficient
to include them into the cross section (\ref{sigN}). The corresponding
``dressed'' diagram is shown in \fig{fig:dress},
where the two exchanged gluons have chains of quark bubbles on them.
Indeed one should sum over connections of each of the gluons to the
quark and the anti-quark in the dipole: this is actually what gives
the factor $\left( 2 - e^{i {\bm l} \cdot {\bm x}} - e^{- i {\bm l}
    \cdot {\bm x}} \right)$ in \eq{sigN}. Since in the eikonal
approximation the gluons light cone momentum is small, $l_- = 0$, the
gluon virtuality is $l^2 = - {\bm l}^2$. The quark bubble chains
introduce two usual denominators into the fixed coupling equation
(\ref{sigN}), turning it into
\begin{align}\label{sig_rc1}
  \sigma^{q{\bar q} N} \, = \, 2 \, \am^2 \, \frac{C_F}{N_c} \, \int
  \frac{d^2 l}{[{\bm l}^2]^2} \, \frac{1}{\left[ 1 + \am \, \beta_2 \,
      \ln \frac{{\bm l}^2}{\mu^2} \right]^2} \, \left( 2 - e^{i {\bm
        l} \cdot {\bm x}} - e^{- i {\bm l} \cdot {\bm x}} \right),
\end{align}
where we have completed the factors of $N_f$ to the full QCD one-loop
beta-function, $N_f \rightarrow - 6 \, \pi \, \beta_2$. \eq{sig_rc1}
obviously contains the physical running coupling squared and can be
rewritten as
\begin{align}\label{sig_rc2}
  \sigma^{q{\bar q} N} \, = \, 2 \, \frac{C_F}{N_c} \, \int \frac{d^2
    l}{[{\bm l}^2]^2} \, \as^2 ({\bm l}^2) \, \left( 2 - e^{i {\bm l}
      \cdot {\bm x}} - e^{- i {\bm l} \cdot {\bm x}} \right).
\end{align}

Fourier-transform of a running coupling constant, as is implied in
\eq{sig_rc2}, is always dangerous due to Landau singularity.  Avoiding
uncertainties introduced by Landau pole we assume that $|\bm x| \ll
1/\Lambda \ll 1/\Lambda_\text{QCD}$ and approximate the integral in
\eq{sig_rc2} by
\begin{align}\label{sig_rc3}
  \sigma^{q{\bar q} N} \, \approx \, 2 \, \frac{C_F}{N_c} \, \pi \,
  {\bm x}^2 \int\limits_\Lambda^{1/|\bm x|} \frac{d l}{l} \, \as^2
  (l^2) \, = \, \, 2 \ \as \left( \frac{1}{{\bm x}^2} \right) \, \as
  \left( \Lambda^2 \right) \, \frac{C_F}{N_c} \, \pi \, {\bm x}^2 \,
  \ln \frac{1}{|\bm x| \, \Lambda}
\end{align}
where we have made use of \eq{ab2}. While indeed in real life $\Lambda
\sim \Lambda_\text{QCD}$, we have made the $\Lambda \gg
\Lambda_\text{QCD}$ assumption to be able to disentangle perturbative
and non-perturbative contributions in the integral of \eq{sig_rc2}.
One should then take the $\Lambda \rightarrow \Lambda_\text{QCD}$
limit to identify the non-perturbative part of the expression. An
analogy would be a finite temperature medium, where $\Lambda \sim T$
and for high temperatures $T \gg \Lambda_\text{QCD}$ our above
approximation would be justified. We would then obtain a factor of
$\as (T^2)$ in \eq{sig_rc3}, which would become a non-perturbative
factor for lower temperatures $T \sim \Lambda_\text{QCD}$.

Comparing \eq{sig_rc3} with \eq{sigN} we see that the inclusion of
running coupling corrections is accomplished by the replacement
\begin{align}
  \am^2 \, \rightarrow \, \as \left( \frac{1}{{\bm x}^2} \right) \,
  \as \left( \Lambda^2 \right).
\end{align}

\eq{N0} with the running coupling corrections included becomes
\begin{align}\label{N0rc}
  N ({\bm x}, {\bm b}) \, = \, 1 - \exp \left[ - \pi \, \as \left(
      \frac{1}{{\bm x}^2} \right) \, \as \left( \Lambda^2 \right) \,
    \frac{C_F}{N_c} \, \rho \, T({\bm b}) \, {\bm x}^2 \, \ln
    \frac{1}{|\bm x| \, \Lambda} \right].
\end{align}

For each multiple rescattering one of the coupling constants is
determined by the non-perturbative scale $\Lambda$ and could be large.
That is why this coupling constant is sometimes absorbed into the
gluon distribution function of the nucleon
\cite{Gribov:1981ac,Ayala:1996em,Ayala:1996ed,AyalaFilho:1997du,Mueller:1989st}.
The other coupling comes with the scale $1/|\bm x|$, which, for
perturbatively small dipoles (corresponding to production of particles
with large transverse momenta) is large, making the corresponding
coupling constant small.



\begin{thebibliography}{10}

\bibitem{Kuraev:1977fs}
E.~A. Kuraev, L.~N. Lipatov, and V.~S. Fadin, {\it {The Pomeranchuk
  singularity in non-Abelian gauge theories}},  {\em Sov. Phys. JETP} {\bf 45}
  (1977) 199--204.

\bibitem{Bal-Lip}
Y.~Y. Balitsky and L.~N. Lipatov {\em Sov. J. Nucl. Phys.} {\bf 28} (1978) 822.

\bibitem{Balitsky:1996ub}
I.~Balitsky, {\it Operator expansion for high-energy scattering},  {\em Nucl.
  Phys.} {\bf B463} (1996) 99--160,
  [\href{http://xxx.lanl.gov/abs/hep-ph/9509348}{{\tt hep-ph/9509348}}].

\bibitem{Balitsky:1997mk}
I.~Balitsky, {\it Operator expansion for diffractive high-energy scattering},
  \href{http://xxx.lanl.gov/abs/hep-ph/9706411}{{\tt hep-ph/9706411}}.

\bibitem{Balitsky:1998ya}
I.~Balitsky, {\it Factorization and high-energy effective action},  {\em Phys.
  Rev.} {\bf D60} (1999) 014020,
  [\href{http://xxx.lanl.gov/abs/hep-ph/9812311}{{\tt hep-ph/9812311}}].

\bibitem{Kovchegov:1999yj}
Y.~V. Kovchegov, {\it Small-x {$F_2$} structure function of a nucleus including
  multiple pomeron exchanges},  {\em Phys. Rev.} {\bf D60} (1999) 034008,
  [\href{http://xxx.lanl.gov/abs/hep-ph/9901281}{{\tt hep-ph/9901281}}].

\bibitem{Kovchegov:1999ua}
Y.~V. Kovchegov, {\it Unitarization of the {BFKL} pomeron on a nucleus},  {\em
  Phys. Rev.} {\bf D61} (2000) 074018,
  [\href{http://xxx.lanl.gov/abs/hep-ph/9905214}{{\tt hep-ph/9905214}}].

\bibitem{Jalilian-Marian:1997jx}
J.~Jalilian-Marian, A.~Kovner, A.~Leonidov, and H.~Weigert, {\it The {BFKL}
  equation from the {Wilson} renormalization group},  {\em Nucl. Phys.} {\bf
  B504} (1997) 415--431, [\href{http://xxx.lanl.gov/abs/hep-ph/9701284}{{\tt
  hep-ph/9701284}}].

\bibitem{Jalilian-Marian:1997gr}
J.~Jalilian-Marian, A.~Kovner, A.~Leonidov, and H.~Weigert, {\it The {Wilson}
  renormalization group for low x physics: Towards the high density regime},
  {\em Phys. Rev.} {\bf D59} (1998) 014014,
  [\href{http://xxx.lanl.gov/abs/hep-ph/9706377}{{\tt hep-ph/9706377}}].

\bibitem{Jalilian-Marian:1997dw}
J.~Jalilian-Marian, A.~Kovner, and H.~Weigert, {\it The {Wilson}
  renormalization group for low x physics: Gluon evolution at finite parton
  density},  {\em Phys. Rev.} {\bf D59} (1998) 014015,
  [\href{http://xxx.lanl.gov/abs/hep-ph/9709432}{{\tt hep-ph/9709432}}].

\bibitem{Jalilian-Marian:1998cb}
J.~Jalilian-Marian, A.~Kovner, A.~Leonidov, and H.~Weigert, {\it Unitarization
  of gluon distribution in the doubly logarithmic regime at high density},
  {\em Phys. Rev.} {\bf D59} (1999) 034007,
  [\href{http://xxx.lanl.gov/abs/hep-ph/9807462}{{\tt hep-ph/9807462}}].

\bibitem{Kovner:2000pt}
A.~Kovner, J.~G. Milhano, and H.~Weigert, {\it Relating different approaches to
  nonlinear {QCD} evolution at finite gluon density},  {\em Phys. Rev.} {\bf
  D62} (2000) 114005, [\href{http://xxx.lanl.gov/abs/hep-ph/0004014}{{\tt
  hep-ph/0004014}}].

\bibitem{Weigert:2000gi}
H.~Weigert, {\it Unitarity at small {B}jorken x},  {\em Nucl. Phys.} {\bf A703}
  (2002) 823--860, [\href{http://xxx.lanl.gov/abs/hep-ph/0004044}{{\tt
  hep-ph/0004044}}].

\bibitem{Iancu:2000hn}
E.~Iancu, A.~Leonidov, and L.~D. McLerran, {\it Nonlinear gluon evolution in
  the color glass condensate. {I}},  {\em Nucl. Phys.} {\bf A692} (2001)
  583--645, [\href{http://xxx.lanl.gov/abs/hep-ph/0011241}{{\tt
  hep-ph/0011241}}].

\bibitem{Ferreiro:2001qy}
E.~Ferreiro, E.~Iancu, A.~Leonidov, and L.~McLerran, {\it Nonlinear gluon
  evolution in the color glass condensate. {II}},  {\em Nucl. Phys.} {\bf A703}
  (2002) 489--538, [\href{http://xxx.lanl.gov/abs/hep-ph/0109115}{{\tt
  hep-ph/0109115}}].

\bibitem{Balitsky:2006wa}
I.~I. Balitsky, {\it {Quark Contribution to the Small-$x$ Evolution of Color
  Dipole}},  {\em Phys. Rev. D} {\bf 75} (2007) 014001,
  [\href{http://xxx.lanl.gov/abs/hep-ph/0609105}{{\tt hep-ph/0609105}}].

\bibitem{Kovchegov:2006vj}
Y.~Kovchegov and H.~Weigert, {\it {Triumvirate of Running Couplings in
  Small-$x$ Evolution}},  {\em Nucl. Phys. {\bf A}} {\bf 784} (2007) 188--226,
  [\href{http://xxx.lanl.gov/abs/hep-ph/0609090}{{\tt hep-ph/0609090}}].

\bibitem{Kovchegov:2006wf}
Y.~V. Kovchegov and H.~Weigert, {\it {Quark loop contribution to BFKL
  evolution: Running coupling and leading-N(f) NLO intercept}},  {\em accepted
  for publication at Nucl. Phys. A} (2006)
  [\href{http://xxx.lanl.gov/abs/hep-ph/0612071}{{\tt hep-ph/0612071}}].

\bibitem{Albacete:2007yr}
J.~L. Albacete and Y.~V. Kovchegov, {\it Solving high energy evolution equation
  including running coupling corrections},  {\em Phys. Rev.} {\bf D75} (2007)
  125021, [\href{http://xxx.lanl.gov/abs/arXiv:0704.0612 [hep-ph]}{{\tt
  arXiv:0704.0612 [hep-ph]}}].

\bibitem{Gribov:1984tu}
L.~V. Gribov, E.~M. Levin, and M.~G. Ryskin, {\it {Semihard Processes in QCD}},
   {\em Phys. Rept.} {\bf 100} (1983) 1--150.

\bibitem{Mueller:1986wy}
A.~H. Mueller and J.-w. Qiu, {\it Gluon recombination and shadowing at small
  values of x},  {\em Nucl. Phys.} {\bf B268} (1986) 427.

\bibitem{McLerran:1994vd}
L.~D. McLerran and R.~Venugopalan, {\it Green's functions in the color field of
  a large nucleus},  {\em Phys. Rev.} {\bf D50} (1994) 2225--2233,
  [\href{http://xxx.lanl.gov/abs/hep-ph/9402335}{{\tt hep-ph/9402335}}].

\bibitem{McLerran:1993ka}
L.~D. McLerran and R.~Venugopalan, {\it Gluon distribution functions for very
  large nuclei at small transverse momentum},  {\em Phys. Rev.} {\bf D49}
  (1994) 3352--3355, [\href{http://xxx.lanl.gov/abs/hep-ph/9311205}{{\tt
  hep-ph/9311205}}].

\bibitem{McLerran:1993ni}
L.~D. McLerran and R.~Venugopalan, {\it Computing quark and gluon distribution
  functions for very large nuclei},  {\em Phys. Rev.} {\bf D49} (1994)
  2233--2241, [\href{http://xxx.lanl.gov/abs/hep-ph/9309289}{{\tt
  hep-ph/9309289}}].

\bibitem{Kovchegov:1996ty}
Y.~V. Kovchegov, {\it Non-abelian {Weizsaecker-Williams} field and a two-
  dimensional effective color charge density for a very large nucleus},  {\em
  Phys. Rev.} {\bf D54} (1996) 5463--5469,
  [\href{http://xxx.lanl.gov/abs/hep-ph/9605446}{{\tt hep-ph/9605446}}].

\bibitem{Kovchegov:1997pc}
Y.~V. Kovchegov, {\it Quantum structure of the non-abelian
  {Weizsaecker-Williams} field for a very large nucleus},  {\em Phys. Rev.}
  {\bf D55} (1997) 5445--5455,
  [\href{http://xxx.lanl.gov/abs/hep-ph/9701229}{{\tt hep-ph/9701229}}].

\bibitem{Jalilian-Marian:1997xn}
J.~Jalilian-Marian, A.~Kovner, L.~D. McLerran, and H.~Weigert, {\it The
  intrinsic glue distribution at very small x},  {\em Phys. Rev.} {\bf D55}
  (1997) 5414--5428, [\href{http://xxx.lanl.gov/abs/hep-ph/9606337}{{\tt
  hep-ph/9606337}}].

\bibitem{Iancu:2003xm}
E.~Iancu and R.~Venugopalan, {\it The color glass condensate and high energy
  scattering in {QCD}},  \href{http://xxx.lanl.gov/abs/hep-ph/0303204}{{\tt
  hep-ph/0303204}}.

\bibitem{Weigert:2005us}
H.~Weigert, {\it Evolution at small {$x_{\text{bj}}$: The Color Glass
  Condensate}},  {\em Prog. Part. Nucl. Phys.} {\bf 55} (2005) 461--565,
  [\href{http://xxx.lanl.gov/abs/hep-ph/0501087}{{\tt hep-ph/0501087}}].

\bibitem{Jalilian-Marian:2005jf}
J.~Jalilian-Marian and Y.~V. Kovchegov, {\it Saturation physics and deuteron
  gold collisions at {RHIC}},  {\em Prog. Part. Nucl. Phys.} {\bf 56} (2006)
  104--231, [\href{http://xxx.lanl.gov/abs/hep-ph/0505052}{{\tt
  hep-ph/0505052}}].

\bibitem{Kovner:1995ja}
A.~Kovner, L.~D. McLerran, and H.~Weigert, {\it Gluon production from
  non{A}belian {W}eizsacker-{W}illiams fields in nucleus-nucleus collisions},
  {\em Phys. Rev.} {\bf D52} (1995) 6231--6237,
  [\href{http://xxx.lanl.gov/abs/hep-ph/9502289}{{\tt hep-ph/9502289}}].

\bibitem{Kovner:1995ts}
A.~Kovner, L.~D. McLerran, and H.~Weigert, {\it Gluon production at high
  transverse momentum in the mclerran-venugopalan model of nuclear structure
  functions},  {\em Phys. Rev.} {\bf D52} (1995) 3809--3814,
  [\href{http://xxx.lanl.gov/abs/hep-ph/9505320}{{\tt hep-ph/9505320}}].

\bibitem{Gyulassy:1997vt}
M.~Gyulassy and L.~D. McLerran, {\it Yang-mills radiation in ultrarelativistic
  nuclear collisions},  {\em Phys. Rev.} {\bf C56} (1997) 2219--2228,
  [\href{http://xxx.lanl.gov/abs/nucl-th/9704034}{{\tt nucl-th/9704034}}].

\bibitem{Kovchegov:1997ke}
Y.~V. Kovchegov and D.~H. Rischke, {\it Classical gluon radiation in
  ultrarelativistic nucleus nucleus collisions},  {\em Phys. Rev.} {\bf C56}
  (1997) 1084--1094, [\href{http://xxx.lanl.gov/abs/hep-ph/9704201}{{\tt
  hep-ph/9704201}}].

\bibitem{Kovchegov:1998bi}
Y.~V. Kovchegov and A.~H. Mueller, {\it Gluon production in current nucleus and
  nucleon nucleus collisions in a quasi-classical approximation},  {\em Nucl.
  Phys.} {\bf B529} (1998) 451--479,
  [\href{http://xxx.lanl.gov/abs/hep-ph/9802440}{{\tt hep-ph/9802440}}].

\bibitem{Kopeliovich:1998nw}
B.~Z. Kopeliovich, A.~V. Tarasov, and A.~Schafer, {\it Bremsstrahlung of a
  quark propagating through a nucleus},  {\em Phys. Rev.} {\bf C59} (1999)
  1609--1619, [\href{http://xxx.lanl.gov/abs/hep-ph/9808378}{{\tt
  hep-ph/9808378}}].

\bibitem{Dumitru:2001ux}
A.~Dumitru and L.~D. McLerran, {\it How protons shatter colored glass},  {\em
  Nucl. Phys.} {\bf A700} (2002) 492--508,
  [\href{http://xxx.lanl.gov/abs/hep-ph/0105268}{{\tt hep-ph/0105268}}].

\bibitem{Kovchegov:2001sc}
Y.~V. Kovchegov and K.~Tuchin, {\it Inclusive gluon production in dis at high
  parton density},  {\em Phys. Rev.} {\bf D65} (2002) 074026,
  [\href{http://xxx.lanl.gov/abs/hep-ph/0111362}{{\tt hep-ph/0111362}}].

\bibitem{Gelis:2002nn}
F.~Gelis and J.~Jalilian-Marian, {\it From dis to proton nucleus collisions in
  the color glass condensate model},  {\em Phys. Rev.} {\bf D67} (2003) 074019,
  [\href{http://xxx.lanl.gov/abs/hep-ph/0211363}{{\tt hep-ph/0211363}}].

\bibitem{Kovchegov:2006qn}
Y.~V. Kovchegov and K.~Tuchin, {\it Production of q anti-q pairs in proton
  nucleus collisions at high energies},
  \href{http://xxx.lanl.gov/abs/hep-ph/0603055}{{\tt hep-ph/0603055}}.

\bibitem{Kovchegov:2005ur}
Y.~V. Kovchegov, {\it Inclusive gluon production in high energy onium onium
  scattering},  {\em Phys. Rev.} {\bf D72} (2005) 094009,
  [\href{http://xxx.lanl.gov/abs/hep-ph/0508276}{{\tt hep-ph/0508276}}].

\bibitem{Dumitru:2001jn}
A.~Dumitru and J.~Jalilian-Marian, {\it Scattering of gluons from the color
  glass condensate},  {\em Phys. Lett.} {\bf B547} (2002) 15--20,
  [\href{http://xxx.lanl.gov/abs/hep-ph/0111357}{{\tt hep-ph/0111357}}].

\bibitem{Dumitru:2002qt}
A.~Dumitru and J.~Jalilian-Marian, {\it Forward quark jets from protons
  shattering the colored glass},  {\em Phys. Rev. Lett.} {\bf 89} (2002)
  022301, [\href{http://xxx.lanl.gov/abs/hep-ph/0204028}{{\tt
  hep-ph/0204028}}].

\bibitem{Kovner:2001vi}
A.~Kovner and U.~A. Wiedemann, {\it {Eikonal evolution and gluon radiation}},
  {\em Phys. Rev.} {\bf D64} (2001) 114002,
  [\href{http://xxx.lanl.gov/abs/hep-ph/0106240}{{\tt hep-ph/0106240}}].

\bibitem{Blaizot:2004wu}
J.~P. Blaizot, F.~Gelis, and R.~Venugopalan, {\it {High energy p A collisions
  in the color glass condensate approach. I: Gluon production and the Cronin
  effect}},  {\em Nucl. Phys.} {\bf A743} (2004) 13--56,
  [\href{http://xxx.lanl.gov/abs/hep-ph/0402256}{{\tt hep-ph/0402256}}].

\bibitem{Marquet:2004xa}
C.~Marquet, {\it {A QCD dipole formalism for forward-gluon production}},  {\em
  Nucl. Phys.} {\bf B705} (2005) 319--338,
  [\href{http://xxx.lanl.gov/abs/hep-ph/0409023}{{\tt hep-ph/0409023}}].

\bibitem{Gelis:2005pt}
F.~Gelis and Y.~Mehtar-Tani, {\it {Gluon propagation inside a high-energy
  nucleus}},  {\em Phys. Rev.} {\bf D73} (2006) 034019,
  [\href{http://xxx.lanl.gov/abs/hep-ph/0512079}{{\tt hep-ph/0512079}}].

\bibitem{Gribov:1981ac}
L.~V. Gribov, E.~M. Levin, and M.~G. Ryskin, {\it Singlet structure function at
  small x: Unitarization of gluon ladders},  {\em Nucl. Phys.} {\bf B188}
  (1981) 555--576.

\bibitem{Catani:1990eg}
S.~Catani, M.~Ciafaloni, and F.~Hautmann, {\it High-energy factorization and
  small x heavy flavor production},  {\em Nucl. Phys.} {\bf B366} (1991)
  135--188.

\bibitem{Kharzeev:2003wz}
D.~Kharzeev, Y.~V. Kovchegov, and K.~Tuchin, {\it Cronin effect and high-p(t)
  suppression in p a collisions},  {\em Phys. Rev.} {\bf D68} (2003) 094013,
  [\href{http://xxx.lanl.gov/abs/hep-ph/0307037}{{\tt hep-ph/0307037}}].

\bibitem{Brodsky:1983gc}
S.~J. Brodsky, G.~P. Lepage, and P.~B. Mackenzie, {\it On the elimination of
  scale ambiguities in perturbative quantum chromodynamics},  {\em Phys. Rev.}
  {\bf D28} (1983) 228.

\bibitem{Mueller:1984vh}
A.~H. Mueller, {\it On the structure of infrared renormalons in physical
  processes at high-energies},  {\em Nucl. Phys.} {\bf B250} (1985) 327.

\bibitem{Parisi:1978bj}
G.~Parisi, {\it Singularities of the borel transform in renormalizable
  theories},  {\em Phys. Lett.} {\bf B76} (1978) 65--66.

\bibitem{David:1983gz}
F.~David, {\it On the ambiguity of composite operators, {I.R.} renormalons and
  the status of the operator product expansion},  {\em Nucl. Phys.} {\bf B234}
  (1984) 237--251.

\bibitem{Zakharov:1992bx}
V.~I. Zakharov, {\it {QCD perturbative expansions in large orders}},  {\em
  Nucl. Phys.} {\bf B385} (1992) 452--480.

\bibitem{Beneke:1998ui}
M.~Beneke, {\it Renormalons},  {\em Phys. Rept.} {\bf 317} (1999) 1--142,
  [\href{http://xxx.lanl.gov/abs/hep-ph/9807443}{{\tt hep-ph/9807443}}].

\bibitem{Beneke:2000kc}
M.~Beneke and V.~M. Braun, {\it Renormalons and power corrections},
  \href{http://xxx.lanl.gov/abs/hep-ph/0010208}{{\tt hep-ph/0010208}}.

\bibitem{Beneke:1994qe}
M.~Beneke and V.~M. Braun, {\it {Naive nonAbelianization and resummation of
  fermion bubble chains}},  {\em Phys. Lett.} {\bf B348} (1995) 513--520,
  [\href{http://xxx.lanl.gov/abs/hep-ph/9411229}{{\tt hep-ph/9411229}}].

\bibitem{Lautrup:1977hs}
B.~Lautrup, {\it On high order estimates in qed},  {\em Phys. Lett.} {\bf B69}
  (1977) 109--111.

\bibitem{'tHooft:1977am}
G.~'t~Hooft, {\it Can we make sense out of 'quantum chromodynamics'?}, .
  Lectures given at Int. School of Subnuclear Physics, Erice, Sicily, Jul 23 -
  Aug 10, 1977.

\bibitem{Balitsky:2008zz}
I.~Balitsky and G.~A. Chirilli, {\it {Next-to-leading order evolution of color
  dipoles}},  {\em Phys. Rev.} {\bf D77} (2008) 014019,
  [\href{http://xxx.lanl.gov/abs/0710.4330}{{\tt 0710.4330}}].

\bibitem{Berger:1996vy}
E.~L. Berger, X.-f. Guo, and J.-w. Qiu, {\it Isolated prompt photon production
  in hadronic final states of $e~+e~-$ annihilation},  {\em Phys. Rev.} {\bf
  D54} (1996) 5470--5495, [\href{http://xxx.lanl.gov/abs/hep-ph/9605324}{{\tt
  hep-ph/9605324}}].

\bibitem{Berger:1995fm}
E.~L. Berger, X.-F. Guo, and J.-W. Qiu, {\it Inclusive prompt photon production
  in hadronic final states of e+ e- annihilation},  {\em Phys. Rev.} {\bf D53}
  (1996) 1124--1141, [\href{http://xxx.lanl.gov/abs/hep-ph/9507428}{{\tt
  hep-ph/9507428}}].

\bibitem{Kovchegov:2005ss}
Y.~V. Kovchegov, {\it {Can thermalization in heavy ion collisions be described
  by QCD diagrams?}},  {\em Nucl. Phys.} {\bf A762} (2005) 298--325,
  [\href{http://xxx.lanl.gov/abs/hep-ph/0503038}{{\tt hep-ph/0503038}}].

\bibitem{Mueller:1989st}
A.~H. Mueller, {\it {Small x Behavior and Parton Saturation: A QCD Model}},
  {\em Nucl. Phys.} {\bf B335} (1990) 115.

\bibitem{Gardi:2006rp}
E.~Gardi, J.~Kuokkanen, K.~Rummukainen, and H.~Weigert, {\it Running coupling
  and power corrections in nonlinear evolution at the high-energy limit},  {\em
  Nucl. Phys.} {\bf A784} (2007) 282--340,
  [\href{http://xxx.lanl.gov/abs/hep-ph/0609087}{{\tt hep-ph/0609087}}].

\bibitem{JalilianMarian:2004da}
J.~Jalilian-Marian and Y.~V. Kovchegov, {\it {Inclusive two-gluon and valence
  quark-gluon production in DIS and p A}},  {\em Phys. Rev.} {\bf D70} (2004)
  114017, [\href{http://xxx.lanl.gov/abs/hep-ph/0405266}{{\tt
  hep-ph/0405266}}].

\bibitem{Chen:1995pa}
Z.~Chen and A.~H. Mueller, {\it {The dipole picture of high-energy scattering,
  the BFKL equation and many gluon compound states}},  {\em Nucl. Phys.} {\bf
  B451} (1995) 579--604.

\bibitem{Lepage:1980fj}
G.~P. Lepage and S.~J. Brodsky, {\it Exclusive processes in perturbative
  quantum chromodynamics},  {\em Phys. Rev.} {\bf D22} (1980) 2157.

\bibitem{Brodsky:1997de}
S.~J. Brodsky, H.-C. Pauli, and S.~S. Pinsky, {\it Quantum chromodynamics and
  other field theories on the light cone},  {\em Phys. Rept.} {\bf 301} (1998)
  299--486, [\href{http://xxx.lanl.gov/abs/hep-ph/9705477}{{\tt
  hep-ph/9705477}}].

\bibitem{Babansky:2002my}
A.~Babansky and I.~Balitsky, {\it Scattering of color dipoles: From low to high
  energies},  {\em Phys. Rev.} {\bf D67} (2003) 054026,
  [\href{http://xxx.lanl.gov/abs/hep-ph/0212075}{{\tt hep-ph/0212075}}].

\bibitem{Kovchegov:1999kx}
Y.~V. Kovchegov and L.~D. McLerran, {\it Diffractive structure function in a
  quasi-classical approximation},  {\em Phys. Rev.} {\bf D60} (1999) 054025,
  [\href{http://xxx.lanl.gov/abs/hep-ph/9903246}{{\tt hep-ph/9903246}}].

\bibitem{Dokshitzer:1977sg}
Y.~L. Dokshitzer, {\it Calculation of the structure functions for deep
  inelastic scattering and e+ e- annihilation by perturbation theory in quantum
  chromodynamics. (in russian)},  {\em Sov. Phys. JETP} {\bf 46} (1977)
  641--653.

\bibitem{Gribov:1972ri}
V.~N. Gribov and L.~N. Lipatov, {\it Deep inelastic e p scattering in
  perturbation theory},  {\em Sov. J. Nucl. Phys.} {\bf 15} (1972) 438--450.

\bibitem{Altarelli:1977zs}
G.~Altarelli and G.~Parisi, {\it Asymptotic freedom in parton language},  {\em
  Nucl. Phys.} {\bf B126} (1977) 298.

\bibitem{Dokshitzer:1993pf}
Y.~L. Dokshitzer and D.~V. Shirkov, {\it On exact account of heavy quark
  thresholds in hard processes},  {\em Z. Phys.} {\bf C67} (1995) 449--458.

\bibitem{Beneke:1995pq}
M.~Beneke and V.~M. Braun, {\it {Power corrections and renormalons in Drell-Yan
  production}},  {\em Nucl. Phys.} {\bf B454} (1995) 253--290,
  [\href{http://xxx.lanl.gov/abs/hep-ph/9506452}{{\tt hep-ph/9506452}}].

\bibitem{Sudakov:1954sw}
V.~V. Sudakov, {\it Vertex parts at very high-energies in quantum
  electrodynamics},  {\em Sov. Phys. JETP} {\bf 3} (1956) 65--71.

\bibitem{Braaten:1993rw}
E.~Braaten and T.~C. Yuan, {\it Gluon fragmentation into heavy quarkonium},
  {\em Phys. Rev. Lett.} {\bf 71} (1993) 1673--1676,
  [\href{http://xxx.lanl.gov/abs/hep-ph/9303205}{{\tt hep-ph/9303205}}].

\bibitem{Kwiecinski:1994zs}
J.~Kwiecinski, A.~D. Martin, P.~J. Sutton, and K.~J. Golec-Biernat, {\it {QCD
  predictions for the transverse energy flow in deep inelastic scattering in
  the HERA small x regime}},  {\em Phys. Rev.} {\bf D50} (1994) 217--225,
  [\href{http://xxx.lanl.gov/abs/hep-ph/9403292}{{\tt hep-ph/9403292}}].

\bibitem{Ayala:1996em}
A.~L. Ayala, M.~B. Gay~Ducati, and E.~M. Levin, {\it Qcd evolution of the gluon
  density in a nucleus},  {\em Nucl. Phys.} {\bf B493} (1997) 305--353,
  [\href{http://xxx.lanl.gov/abs/hep-ph/9604383}{{\tt hep-ph/9604383}}].

\bibitem{Ayala:1996ed}
F.~Ayala, A.~L., M.~B. Gay~Ducati, and E.~M. Levin, {\it Unitarity boundary for
  deep inelastic structure functions},  {\em Phys. Lett.} {\bf B388} (1996)
  188--196, [\href{http://xxx.lanl.gov/abs/hep-ph/9607210}{{\tt
  hep-ph/9607210}}].

\bibitem{AyalaFilho:1997du}
A.~L. Ayala~Filho, M.~B. Gay~Ducati, and E.~M. Levin, {\it Parton densities in
  a nucleon},  {\em Nucl. Phys.} {\bf B511} (1998) 355--395,
  [\href{http://xxx.lanl.gov/abs/hep-ph/9706347}{{\tt hep-ph/9706347}}].

\end{thebibliography}

\providecommand{\href}[2]{#2}\begingroup\raggedright\endgroup

\end{document}